\documentclass[sn-mathphys-num,referee]{sn-jnl}
\usepackage{graphicx}%
\usepackage{multirow,ulem}%
\usepackage{amsmath,amssymb,amsfonts}%
\usepackage{amsthm}%
\usepackage{mathrsfs}%
\usepackage[title]{appendix}%
\usepackage{xcolor}%
\usepackage{textcomp}%
\usepackage{manyfoot}%
\usepackage{booktabs}%
\usepackage{algorithm}%
\usepackage{algorithmicx}%
\usepackage{algpseudocode}%
\usepackage{listings}%

\begin{document}

\title[Geometric indicators of local plasticity in glasses measured by scanning small-beam diffraction]{Geometric indicators of local plasticity in glasses measured by scanning small-beam diffraction}

\author*[1]{\fnm{Amelia C. Y.} \sur{Liu}}\email{amelia.liu@monash.edu}
\author[1]{\fnm{Huyen} \sur{Pham}}\email{huyen.pham@monash.edu}

\author[2]{\fnm{Arabinda} \sur{Bera}}\email{arabinda.bera@unimi.it}

\author[3,1]{\fnm{Timothy C.} \sur{Petersen}}\email{timothy.petersen@monash.edu}

\author[4]{\fnm{Timothy W.} \sur{Sirk}}\email{timothy.w.sirk.civ@army.mil}

\author[5]{\fnm{Stephen T.} \sur{Mudie}}\email{stephenm@ansto.gov.au}

\author[6]{\fnm{Rico F.} \sur{Tabor}}\email{rico.tabor@monash.edu}

\author[7]{\fnm{Juan} \sur{Nunez-Iglesias}}\email{juan.nunez-iglesias@monash.edu}

\author*[2]{\fnm{Alessio} \sur{Zaccone}}\email{alessio.zaccone@unimi.it}

\author*[8,9]{\fnm{Matteo} \sur{Baggioli}}\email{b.matteo@sjtu.edu.cn}

\affil*[1]{\orgdiv{School of Physics and Astronomy}, \orgname{Monash University}, \orgaddress{\street{Wellington Road}, \city{Clayton}, \postcode{3800}, \state{Victoria}, \country{Australia}}}

\affil[2]{\orgdiv{Department of Physics ``A. Pontremoli"}, \orgname{University of Milan}, \orgaddress{\street{via Celoria 16}, \city{Milan}, \postcode{20133}, \state{Lombardy}, \country{Italy}}}

\affil[3]{\orgdiv{Monash Centre for Electron Microscopy}, \orgname{Monash University}, \orgaddress{\street{Wellington Road}, \city{Clayton}, \postcode{3800}, \state{Victoria}, \country{Australia}}}

\affil[4]{\orgdiv{Polymers Branch}, \orgname{US DEVCOM Army Research Laboratory}, \city{Aberdeen Proving Ground}, \postcode{21005}, \state{Maryland}, \country{USA}}

\affil[5]{\orgdiv{Australian Synchrotron}, \orgname{ANSTO}, \orgaddress{\street{Blackburn Road}, \city{Clayton}, \postcode{3168}, \state{Victoria}, \country{Australia}}}

\affil[6]{\orgdiv{School of Chemistry}, \orgname{Monash University}, \orgaddress{\street{Wellington Road}, \city{Clayton}, \postcode{3800}, \state{Victoria}, \country{Australia}}}

\affil[7]{\orgdiv{Monash eResearch Centre}, \orgname{Monash University}, \orgaddress{\street{Wellington Road}, \city{Clayton}, \postcode{3800}, \state{Victoria}, \country{Australia}}}

\affil[8]{\orgdiv{Wilczek Quantum Center, School of Physics and Astronomy}, \orgname{Shanghai Jiao Tong University}, \orgaddress{\street{Dongchuan Road}, \city{Shanghai}, \postcode{200240}, \state{Jiangsu}, \country{China}}}

\affil[9]{\orgdiv{Shanghai Research Center for Quantum Sciences}, \orgname{Shanghai Jiao Tong University}, \orgaddress{\street{Dongchuan Road}, \city{Shanghai }, \postcode{201315}, \state{Jiangsu}, \country{China}}}

\abstract{The notion of defects in crystalline phases of matter has been extremely powerful for understanding crystal growth, deformation and melting. Many of these discontinuities in the periodic order of crystals are well described by the Burgers vector, derived from the particle displacements, which encapsulates the direction and magnitude of slip relative to the undeformed state. Since the reference structure of the crystal is known \textit{a priori}, the Burgers vector can be determined experimentally using both imaging and diffraction methods to measure the final lattice distortion, and thus infer the particle displacements. Glasses have structures that lack the periodicity of crystals, and thus a well-defined reference state. Yet, measurable structural parameters can still be obtained from diffraction from a glass. Here we examine the usefulness of these parameters to probe deformation in glasses. We find that co-ordinated transformations in the centrosymmetry of local particle arrangements are a strong marker of plastic events. For a glass, determining the local distortions corresponding to these plastic events requires measurements before and after deformation. We investigate two geometric indicators that can be derived from these distortions, namely the continuous Burgers vector and the quadrupolar strain. We find that the Burgers vector again emerges as a robust and sensitive metric for understanding local structural transformations due to mechanical deformation, even in disordered glasses.}  

\maketitle

\section{Introduction}\label{sec1}
The understanding that dislocations are the carriers of plasticity in crystals has led to many advances in the engineering of alloys and semiconductors. The cores of these defects are highly localized, occurring along a line where the periodic order is discontinuous and the strain diverges. Away from the core, the lattice distortion decays \cite{cottrell1961dislocations}. These crystalline defects are well-characterized by the Burgers vector that reveals the magnitude and direction of the material slip that produced the defect \cite{burgers1}. Formally, the Burgers vector is calculated from the closed contour integral over increments in the particle displacement field \cite{kleinert1989gauge,mura}, 
\begin{equation}
    b_{i} = -  \oint\, du_{i} =  -\oint\, \frac{\partial u_{i}}{\partial x_{j}} dx_{j}
    \label{burgers}
\end{equation}
Here $b_{i}$ is the Burgers vector, $u_{i}$ is a particle displacement and $x_{j}$ is a Cartesian coordinate. The indices $i,j = 1,2,3$ refer to different Cartesian directions. $\frac{\partial u_{i}}{\partial x_{j}}$ form the components of the deformation or distortion tensor that transforms the structure to the deformed state \cite{nyebook} (see Methods~\ref{calcparam}). By Stokes theorem, the Burgers integral is non-zero where the strain compatibility conditions do not hold, or where the strain is not compatible with a single-valued displacement field \cite{mura,kleinert1989gauge}.

The Burgers vector points in the direction of the slip plane and its magnitude is a ``quantized'' fraction of the lattice constant of the crystal \cite{10.1093/oso/9780198851875.001.0001}, completely describing the deformation. For a crystal, the undeformed, reference configuration may be assumed, and thus the particle displacements in Eqn.~\ref{burgers} can be inferred from the deformed structure alone. Thus for crystals, the Burgers vector can be measured from direct imaging of lattice planes or atomic columns, for example in scanning/transmission electron microscopy \cite{HYTCH1998131} or from electron and x-ray diffraction-based techniques \cite{wheeler,cloete}. The description and detection of plastic defects in amorphous materials is not as straightforward as for crystals and remains an area of active current research. In the following discussion we highlight some aspects of this subject, especially those that link to parameters we measure with small-beam diffraction, and also note some comprehensive summaries of recent progress \cite{zaccone2023theory,RevModPhys.90.045006,PhysRevMaterials.4.113609,falk_langer,berthier2025yielding}.

Several works have attempted to broaden the concept of defects in crystals to continuous defects appropriate to describe amorphous solids and glasses \cite{RevModPhys.80.61,Steinhardt1979}. Examining Eqn.~\ref{burgers}, there is no special difficulty with this, if the particle displacements are known. In the most general sense, a non-zero Burgers vector detects strain incompatibility in a region due to displacement and rotation fields \cite{kleinert1989gauge}. In a crystalline lattice, the initial and distorted structural configurations possess inversion symmetry or centrosymmetry in the arrangement of nearest neighbours. The force balance in both the initial and distorted configurations means that particle displacements are \textit{affine}, or proportional to the applied distortion. In contrast, for glasses that lack this local centrosymmetry, the forces from neighbours after distortion are unbalanced and additional, \textit{non-affine} displacements occur \cite{zaccone2023theory}. The Burgers vector magnitude calculated from particle displacements in a glass strongly correlates with plasticity \cite{baggioli} and interestingly, it is only the non-affine displacements that contribute to the Burgers vector (Eqn.~\ref{burgers}) \cite{PhysRevE.105.024602}.

Other researchers, inspired by analogies with electrostatics and the classic Eshelby ellipsoidal inclusion problem in mechanics \cite{eshelby}, have identified the local quadrupolar strain field as the lowest order elastic multipole that is not forbidden by geometric conservation laws \cite{Kumar_2024,PhysRevE.104.024904}. The quadrupolar strain field $Q_{ij}$ can be calculated from the symmetric part of the distortion tensor, or strain, $\epsilon_{ij}$. The non-affine strain is decomposed into a trace ($\mathrm{Tr}$) and a traceless component:
\begin{equation}
    \epsilon_{ij} = \frac{1}{2}  \mathrm{Tr}(\epsilon_{ij}) I + Q_{ij}  .
    \label{quad}
\end{equation}
Here $I$ is the identity tensor and $Q_{ij}$ is the quadrupolar field which has a magnitude and direction (Methods~\ref{calcparam}) that quantifies the deviatoric component of the strain that is a pure shear distortion without any volume change \cite{Asaro_Lubarda_2006}. There have been several experimental observations of quadrupolar strain fields in colloidal \cite{jensen2014local} and metallic glasses \cite{kang2023direct}.

More recently, several articles have focused on vortex and antivortex defects in the direction of the non-affine displacement field and vibrational eigenvectors \cite{desmarchelier2024topological, wu2023, bera2024soft,vaibhav2024experimental}. Here, vortex defects are found in areas in which a closed contour integral over the gradient in the vector direction equals an integer multiple of $2\pi$, defining the \textit{winding number} $\mathcal{W}$. Softness and plasticity seem to correlate particularly with the $\mathcal{W}=-1$ topological defects \cite{wu2023}, which resemble the quadrupolar displacement fields typical of an Eshelby-type event \cite{desmarchelier2024topological}. This analysis can be also extended to 3D systems by generalizing the concept of winding number \cite{bera2024hedgehog}. Currently, the relationships between all these defect descriptors are not known, but being actively researched \cite{Baggioli2023} to find the most fundamental and powerful descriptor of the structural transformations that mediate deformation in amorphous materials and place the mechanics of disordered solids on the same sure footing as for crystals.

A key component of understanding plasticity in amorphous materials is predicting which structural configurations will transform under an applied load \cite{PhysRevMaterials.4.113609}. Structural centrosymmetry has emerged as a critical local parameter to predict the elastic response of an amorphous material \cite{lemaitre2006sum,maloney2006amorphous,milkus2016local,zaccone2011approximate,schlegel2016local,zaccone2023theory}.
The force acting on a given particle, $\alpha$, after affine deformation, $\gamma$, is given by $\underline{f}_{\alpha} = \underline{\Xi}_{\alpha}\gamma$. Here $\underline{\Xi}_{\alpha,i j}$ is a vector quantity (denoted by an underline) that quantifies the magnitude and direction of the asymmetry in the arrangements of the $\beta$ particles around a central particle, $\alpha$. $\underline{\Xi}_{\alpha,i j}$ is only zero for a perfect centrosymmetric lattice. Thus, the expression for $\underline{f}_{\alpha}$ predicts additional, non-affine displacements in non-centrosymmetric crystals and amorphous materials to equilibrate a non-zero force after the affine deformation. The Greek letters $\alpha,\beta$ refer to particles while Latin letters $i,j$ represent the Cartesian directions and summation over $i,j$ is assumed. Please note the difference to the index notation employed previously \cite{lemaitre2006sum,maloney2006amorphous,milkus2016local,zaccone2011approximate,schlegel2016local,zaccone2023theory} to maintain consistency with the notation employed in the equations defining the Burgers vector and quadrupole (Equations~\ref{burgers} and ~\ref{quad}). $\underline{\Xi}_{\alpha,i j} = -\kappa R_{0} \sum_{\beta} \hat{n}_{\alpha \beta, i} \hat{n}_{\alpha \beta, j} \underline{\hat{n}}_{\alpha \beta}$, where $\kappa$ is the spring constant and anharmonic bond-tension terms are neglected. Hence, for a shear deformation, $ij$ refer to the shear plane directions and are fixed in the above expression. $\hat{n}_{\alpha \beta}$ is the unit vector pointing from particle $\beta$ to $\alpha$ and $\hat{n}_{\alpha \beta, i}$ is the i-th component of this.

$\underline{\Xi}_{\alpha}$ can also be formulated into a normalized, scalar, order parameter $F_{IS}$ where $IS$ denotes inversion symmetry, or centrosymmetry. This order parameter can have values between $0$ and $1$ for arrangements that show maximum permissible inversion symmetry-breaking and perfect inversion symmetry, respectively \cite{milkus2016local,schlegel2016local,zaccone2023theory}.
This parameter is defined by \cite{milkus2016local,schlegel2016local,zaccone2023theory}:\begin{equation}
    F_{IS} = 1 - \frac{\sum_{ij}\mathopen|\underline{\Xi}_{ij}\mathclose|^{2}}{\sum_{ij}\mathopen|\underline{\Xi}_{ij}\mathclose|_{ISB}^{2}}.
    \label{centro}
\end{equation} 
Here, $\mathopen|\underline{\Xi}_{ij}\mathclose|_{ISB}^{2} = \kappa^{2} R_{0}^{2} \sum_{\alpha \beta} (\hat{n}_{\alpha \beta,i} \hat{n}_{\alpha \beta,j})^{2}$ is the value for $\mathopen|\underline{\Xi}_{ij}\mathclose|^{2}$ when there is maximum breaking of the inversion, or centrosymmetry \cite{milkus2016local,zaccone2023theory} and $ISB$ denotes inversion-symmetry breaking \cite{milkus2016local}.

Advances in beam definition and detector technologies have opened up many new opportunities for spatially resolved, small, coherent-beam diffraction experiments, well-suited to solving problems in complex materials where disorder plays a role \cite{Bøjesen_2020}. Small, coherent-beam diffraction methods from disordered materials produce ``speckle" diffraction patterns. Many interesting parameters have been proposed that quantify different aspects of the local structure and dynamics by measuring fluctuations in the diffracted intensity in space and time \cite{treacy2005fluctuation, zhang2018spatially,hurley1997resolving} or angular correlations \cite{howie1985intensity,gibson2010substantial,wochner2009x,altarelli2010x,PhysRevLett.110.205505,martin2017orientational} or strain \cite{pekin,doi:10.1126/sciadv.abn0681}. Such diffraction-based approaches are especially valuable for experimental disordered systems where you cannot spatially resolve particle positions and displacements in three dimensions, for example sub-micron colloidal glasses, molecular and atomic glasses. They are highly sensitive to small changes in structural parameters, spatially resolving changes in local structure and stability with differences $<1\%$ \cite{doi:10.1126/sciadv.abn0681}.  
\begin{figure*}[!htbp]
\centering
\includegraphics[width=0.95\linewidth]{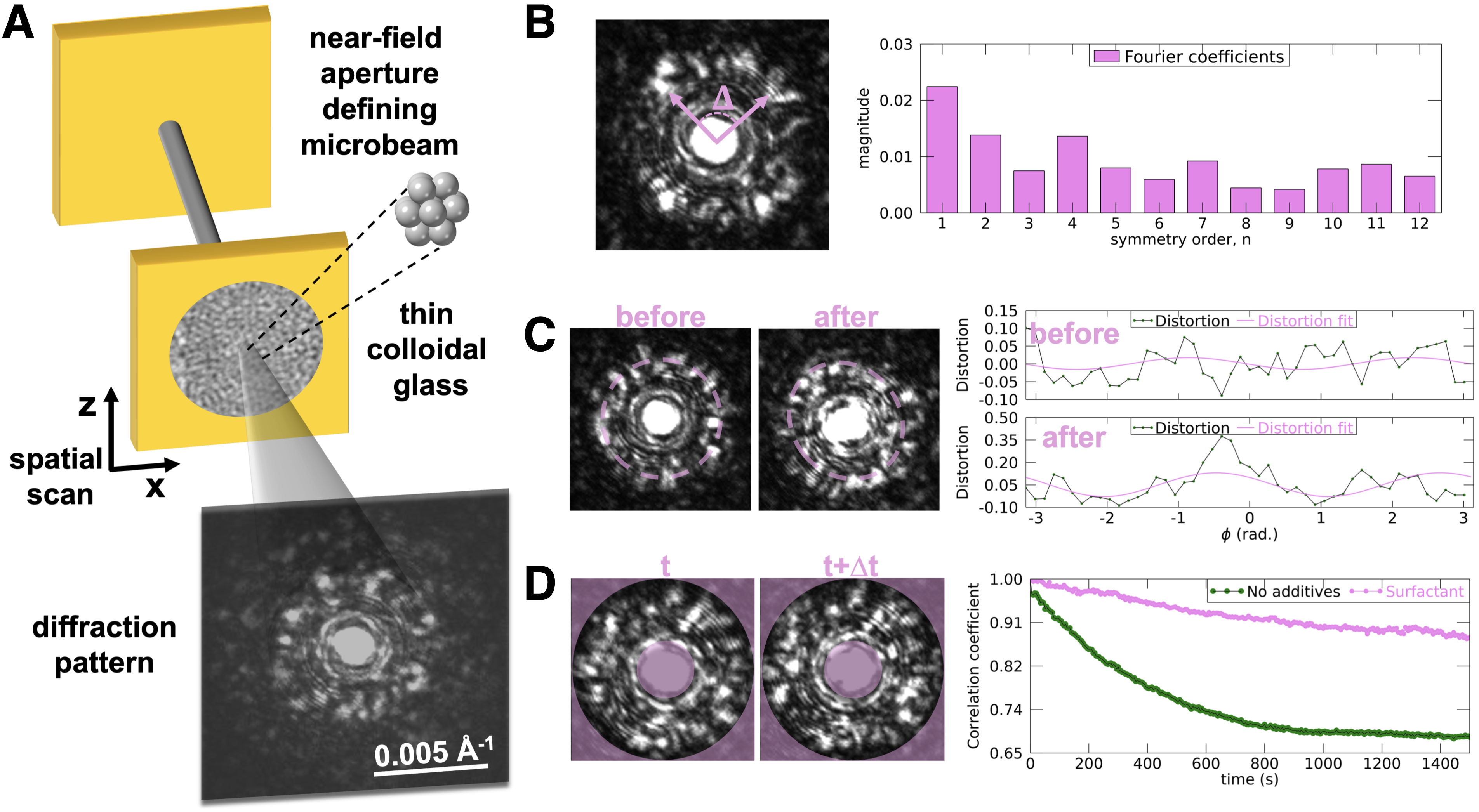}
\caption{Scanning micro-beam x-ray scattering (A) A thin colloidal glass is scanned under a micro-x-ray beam defined by a near-field aperture. Many local structural parameters can be measured from the diffraction patterns such as (B) degree of centrosymmetry in the plane of the specimen from examining angular correlations at angular separations of $\Delta$ (C) strain or distortion (D) stability from cross-correlations at different times, separated by $\Delta t$.}
\label{fig1a}
\end{figure*}
 
 A recent experimental scanning micro-small angle x-ray scattering study measured three local structural parameters in a colloidal glass that showed significant changes with applied deformation \cite{doi:10.1126/sciadv.abn0681}. In this experiment, a thin ($\approx 20~\mu \mathrm{m}$) colloidal glass composed of $300~\mathrm{nm}$ diameter particles is scanned under a micro-x-ray beam defined by a near-field aperture and tuned to the size of a co-ordination polyhedron in the glass as shown in Figure~\ref{fig1a} (A). Several local parameters can be measured simultaneously from the resultant diffraction patterns as shown in the figure: (B) the local centrosymmetry from the breakdown in Friedel symmetry and appearance of odd angular symmetries in the diffraction pattern  \cite{acyliu2015,doi:10.1126/sciadv.abn0681}, (C) the local strain from an ellipse fitted to the intensity centre-of-mass along different radial directions and (D) the local stability from cross-correlations between diffracted intensities in the first diffracted ring at different times \cite{doi:10.1126/sciadv.abn0681}. 
 
 This article provides further confirmation that these local parameters give microscopic insight into deformation mechanisms in a glass using calculations from a simulated polymer glass under simple shear. We further extend the capabilities of this experimental approach by demonstrating via simulation how the distortion tensor can be measured in a glass where, in contrast to crystals, the initial structural configuration is not known, and so the particle displacements can't be inferred from the final structure alone. Thus, in a glass, maps of the local distortions before and after a deformation increment has been applied are required to measure the components of the distortion tensor. The distortion tensor allows us to access well-defined measures of plastic structural rearrangements in a glass, namely the Burgers vector and the quadrupolar field. We find that the Burgers vector and quadrupolar field both concentrate in areas where the non-affine displacement field is large, indicating a plastic event. Yet, despite this, for several reasons, the Burgers vector emerges as capturing unique and essential aspects of the material's response to deformation, like direction and length scale of the coordinated slips. These plastic events are accompanied by changes in the local centrosymmetry of particle arrangements enabled by out-of-plane motion of individual particles. We show how these parameters can be measured in a scanning micro-beam small-angle x-ray scattering experiment of a colloidal glass, demonstrating a pathway towards connecting structural defects and geometry that has been very illuminating for crystals, but in this case for a glass \cite{doi:10.1080/00018730903043166}.
 
\section{Continuous Burgers vector in a simulated glass}\label{sec2} 
\begin{figure}[!htbp]
\centering
\includegraphics[width=0.9\linewidth]{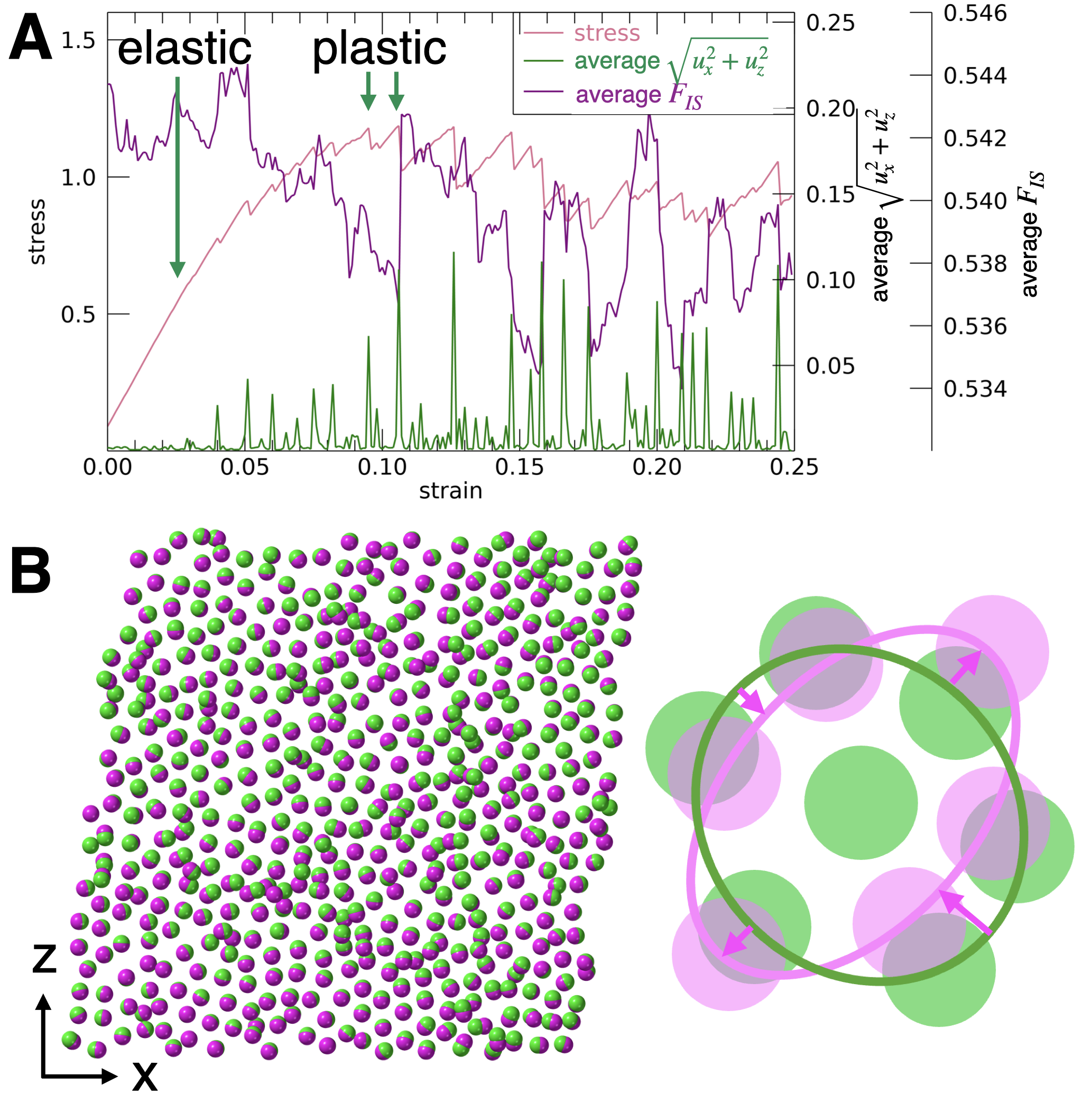}
\caption{Model glass undergoing simple shear. (A) Stress-strain curve (pink line) for the range of strain values studied $0.00 \leq \gamma \leq 0.25$. Green arrows show the values of strain in the elastic and plastic regimes for the configurations studied in Figure~\ref{fig2}. Green and purple lines overlaid show the average magnitudes of the displacements ($\sqrt{u_{x}^2+u_{z}^2}$) and centrosymmetry ($F_{IS}$) from slices in the x-z plane, respectively. (B) Left - Particle positions in a slice of the model before (green) and after (pink) a strain increment is applied that induce distortion of nearest-neighbor configurations illustrated - Right.}
\label{fig1}
\end{figure}
Initially we simulate the measurable parameters from a model bead-spring (Kremer-Grest) polymer glass (with average interparticle distance of $1$ scaled unit) that undergoes simple shear up to a value of $\gamma = 0.25$ (preparation details in \cite{baggioli} and Methods~\ref{model}). We calculate the in-plane (x-z) and out-of-plane (y) non-affine displacements in slices of the three-dimensional model that are one nearest-neighbor polyhedral diameter thick and interpolate the fields (see \cite{baggioli,bera2024soft} and Methods~\ref{model}). The distortion tensor components and derived parameters such as the Burgers vector and quadrupolar strain field are calculated numerically from the interpolated displacement fields in the (x-z) plane. We also calculate the local structural centrosymmetry of the polyhedra in the plane of our slice (Methods~\ref{calcparam}). This quasi-2D approach is comparable to structural projection in a transmission diffraction geometry experiment with the plane of the applied strain perpendicular to the beam direction.

In Figure~\ref{fig1} (A) we show the average magnitudes of the non-affine displacement and centrosymmetry in the plane of the applied strain (x-z) at each strain increment. The non-affine displacement magnitudes increase sharply at the serrations in the stress-strain curve. We associate these sharp stress-drops with plastic events although there are many further considerations regarding their irreversibility \cite{PhysRevE.82.066116} and how the stress-strain curve can be reproduced by localized plastic events \cite{desmarchelier2024topological}. The local centrosymmetry has a unique contrasting behaviour to the non-affine displacement magnitude, generally decreasing just before a serration, and then increasing again straight after, but having an overall downward trend. Physically, this trend may correspond to structural softening just prior to a plastic event, and then rigidity as the system finds a new stable configuration after the stress drop. The overall decrease in local centrosymmetry with deformation has been measured previously experimentally \cite{doi:10.1126/sciadv.abn0681} reflecting a less equilibrated structure after deformation. The Burgers magnitude, quadrupolar magnitude and magnitude of the change in centrosymmetry also show sharp, positive jumps at plastic events (Supplementary Material Figure~S.1).  Figure~\ref{fig1} (B) shows the particle positions in a slice of the model before (green) and after (pink) a strain increment is applied, showing great inhomogeneity in the local distortions. 

\begin{figure*}[!htbp]
\centering
\includegraphics[width=\linewidth]{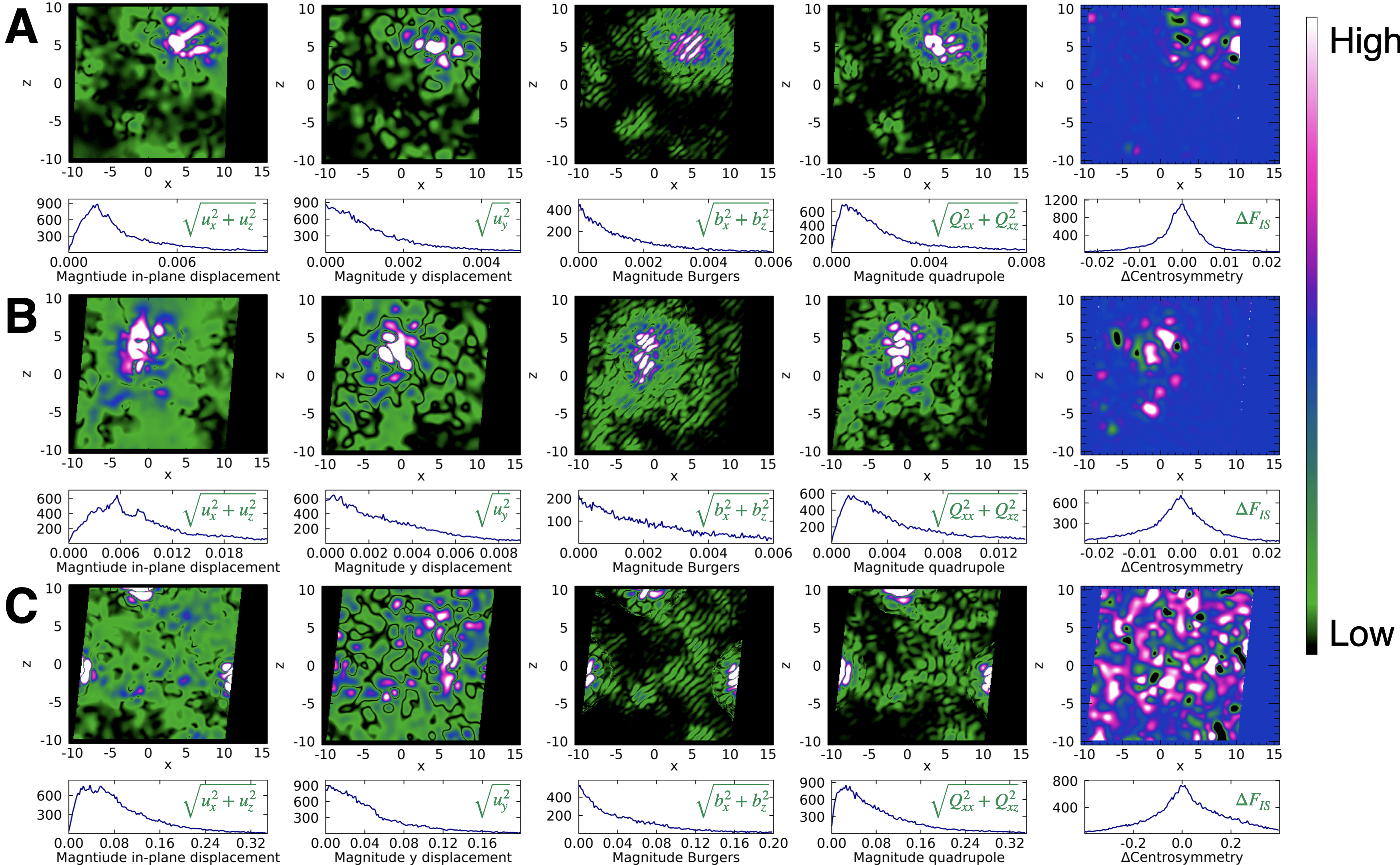}
\caption{In-plane ($\sqrt{u_{x}^2+u_{z}^2}$) and out-of-plane ($\sqrt{u_{y}^2}$) non-affine displacement magnitudes, Burgers vector magnitude ($\sqrt{b_{x}^2+b_{z}^2}$), quadrupolar magnitude ($\sqrt{Q_{xx}^2+Q_{xz}^2}$) and change in local structural centrosymmetry ($\Delta F_{IS}$) for a x-z slice of the glass simulation at  (A) $\gamma = 0.025$ (B) $\gamma = 0.093$ (C) $\gamma = 0.108$ (see green arrows overlaid on the stress-strain curve of Figure~\ref{fig1} (A)). Histograms of the mapped parameters are shown below each map. Note the order of magnitude increase in all the parameter values at the system-spanning plastic event shown in (C).}
\label{fig2}
\end{figure*}

To investigate this, we spatially map these parameter magnitudes for one slice of the simulation cell in Figure~\ref{fig2} with histograms of the mapped values given below each map. We examine the spatial distribution of parameters in the elastic regime ($\gamma = 0.025$), at a larger plastic event ($\gamma = 0.093$) where the stress-strain curve is departing from linearity and a very large plastic event close to the failure point ($\gamma = 0.108$) with these points marked by green arrows in the stress-strain curve in Figure~\ref{fig1} (A). For all the plastic events it is clear that the in-plane (x-z) and out-of-plane (y) displacements correlate strongly with the Burgers vector and quadrupolar strain field and regions where the change in the structural centrosymmetry is large (both negative and positive). The lateral size of the plastic event in the maps correlates with the parameter magnitudes seen in the histograms. Going from isolated and small, to large, and system spanning, the magnitudes increase by a factor of ten. 

\begin{figure}[!htbp]
\centering
\includegraphics[width=\linewidth]{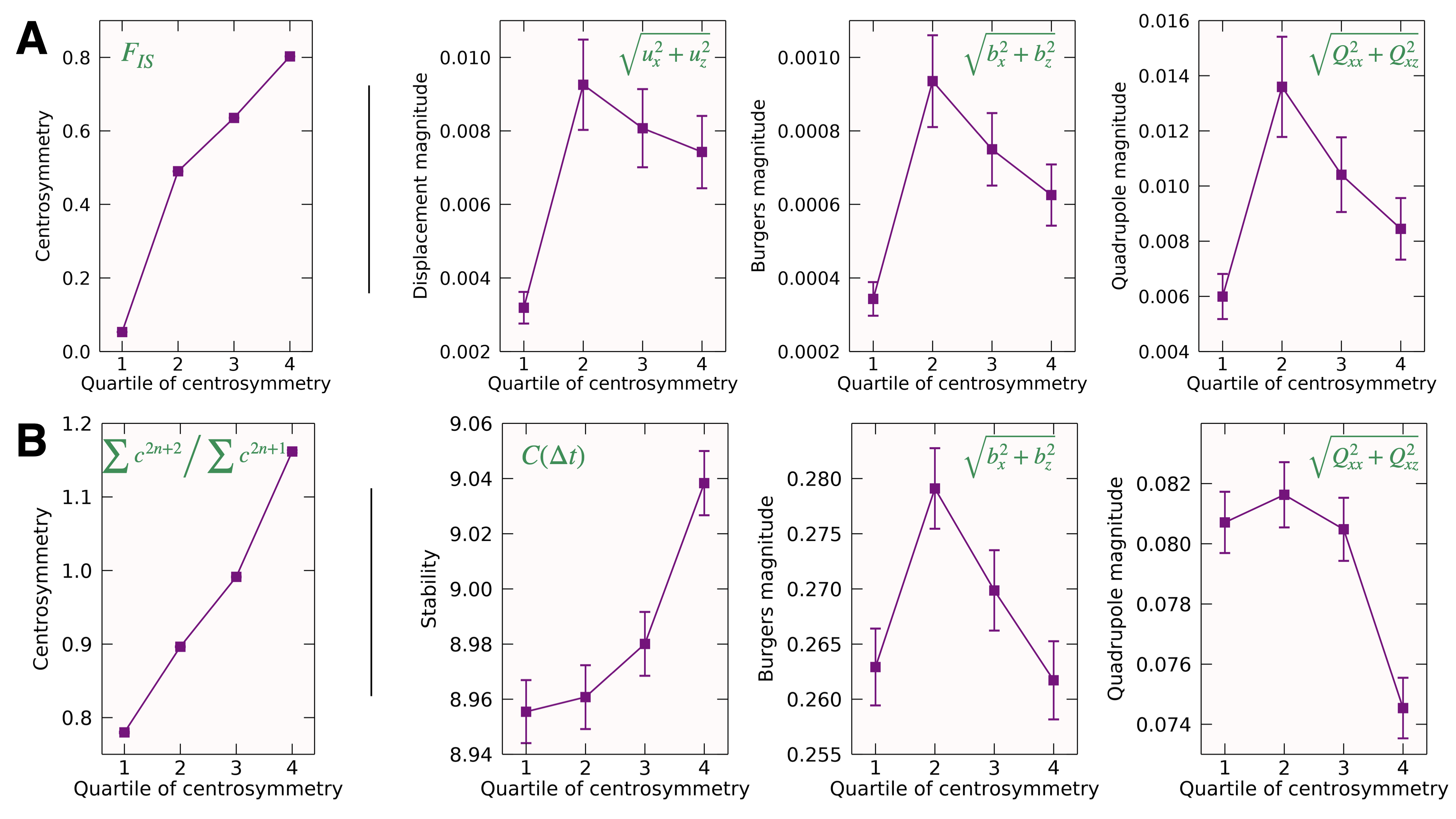}
\caption{(A) Average values and standard error of localized parameters from simulation (centrosymmetry $F_{IS}$, displacement magnitude $\sqrt{u_{x}^2+u_{z}^2}$, Burgers magnitude $\sqrt{b_{x}^2+b_{z}^2}$ and quadrupole magnitude $\sqrt{Q_{xx}^2+Q_{xz}^2}$) that correspond to the different quartiles in the value of the centrosymmetry in the configuration prior to the application of a strain increment. These values were averaged over the whole simulation volume (sliced into twenty x-z slices) and all strain increments. (B) Average values and standard error of localized parameters from experiment (centrosymmetry $\sum c^{2n+2}\slash \sum c^{2n+1}$, stability $C(\Delta t)$ , Burgers magnitude $\sqrt{b_{x}^2+b_{z}^2}$ and quadrupole magnitude $\sqrt{Q_{xx}^2+Q_{xz}^2}$) that correspond to the different quartiles in the value of the centrosymmetry prior to the application of strain. Values were determined from raw experimental maps of each parameter (Methods~\ref{expt}). The magnitude of the local centrosymmetry in different quartiles (shown to the left of the black lines in (A) and (B)) predicts the values of the other parameters.}
\label{fig3}
\end{figure}

\begin{figure}[!htbp]
\centering
\includegraphics[width=\linewidth]{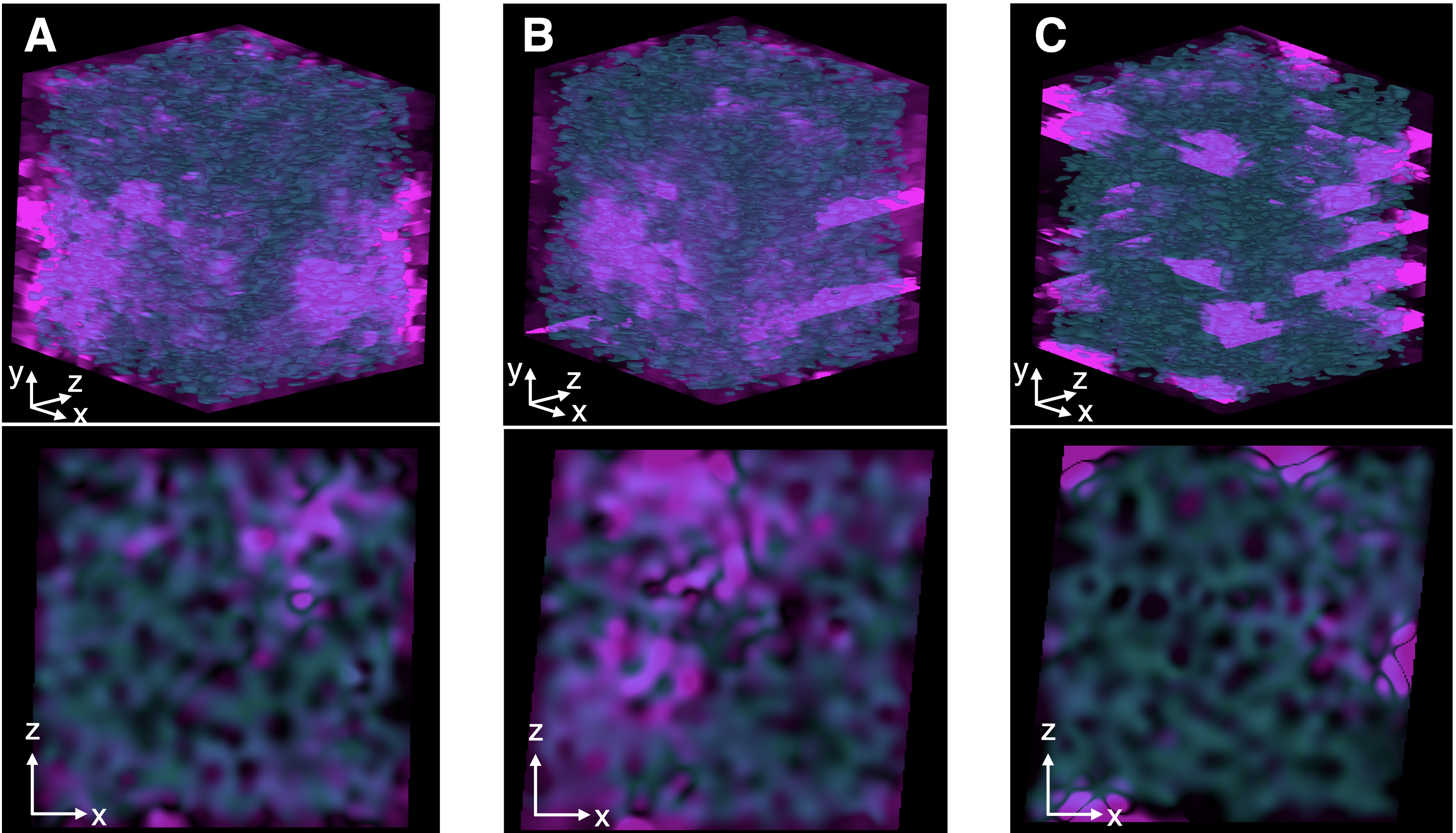}
\caption{Visualizations of the local structural centrosymmetry (green - as an isosurface with a value of $F_{IS} = 0.75$) at $\gamma - \delta \gamma$ and the non-affine in-plane displacement field (pink) at $\gamma$ for (A) $\gamma = 0.025$ (B) $\gamma = 0.093$ (C) $\gamma = 0.108$ (see green arrows overlaid on the stress-strain curve of Figure~1 from the main text). These parameters are displayed as a volume and also as a slice from the middle of the volume. The local structural centrosymmetry forms a stabilizing network that is complementary to the magnitude of the non-affine displacements. Note the order of magnitude increase in the non-affine displacements in (C). Settings for the visualizations are detailed in Methods~\ref{calcparam}.}
\label{fig4}
\end{figure}

Previous experimental work found that local centrosymmetry strongly predicted local structural stability \cite{doi:10.1126/sciadv.abn0681} confirming theory \cite{milkus2016local,schlegel2016local,zaccone2023theory}. We see a strikingly similar relationship in our strained glass models as shown in Figure~\ref{fig3} which displays the average and standard deviation of parameter values that correspond to different quartiles in the centrosymmetry of the configuration prior to the application of a strain increment. Clearly, higher values of local structural centrosymmetry in the configuration at $\gamma -\delta \gamma$ correspond to lower magnitudes of the non-affine displacements induced by the strain increment (at $\gamma$). Interestingly, this trend only holds down to a lower threshold value in centrosymmetry, potentially corresponding to under-coordinated environments where the net force on particles is too low to trigger a rearrangement. Recent work has illuminated the complexity in the statistics of localized structural parameters (in this case local yield stress \cite{PhysRevE.97.033001}) to predict plasticity. These include the direction of applied strain and effects of neighboring environments that can modify the strength of the relationship between the local parameter and plastic rearrangement. In Figure~\ref{fig4} we show volumetric renderings at the same strain steps as Figure~\ref{fig2}. These images show how the structural centrosymmetry (green isosurface) is a network within the volume and that focal points in the displacement magnitude (pink) occur in regions where this centrosymmetry is low.

The Burgers and quadrupolar magnitudes follow the same trends as the displacement as seen in Figure~\ref{fig3}. This demonstrates that the Burgers vector and quadrupolar component of the strain field are strong indicators of structural rearrangements and plasticity even though they reflect slightly different aspects of the deformation eg strain incompatibility and deviatoric strain.

\begin{figure}[!htbp]
\centering
\includegraphics[width=0.8\linewidth]{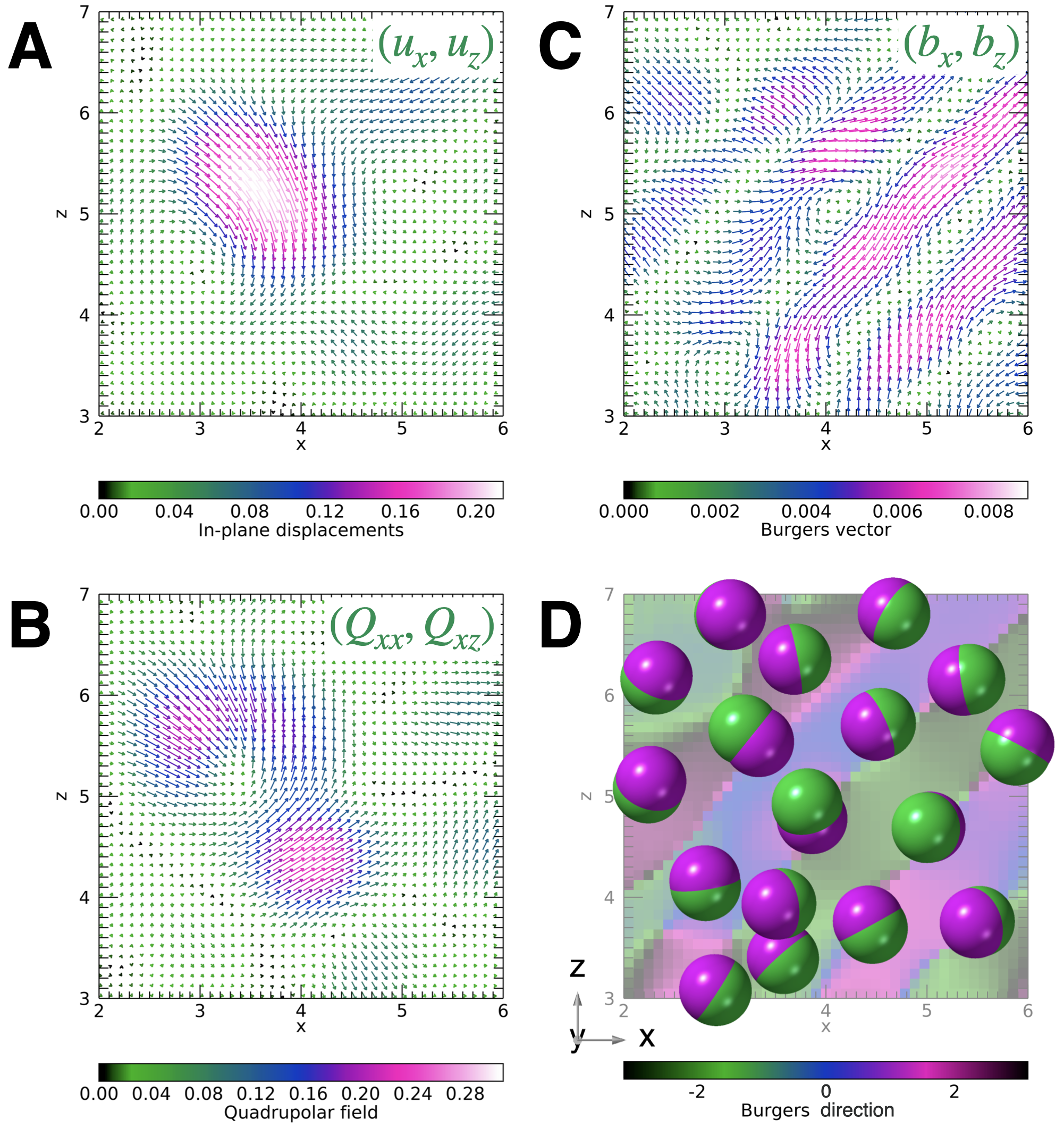}
\caption{Interpolated fields for a slice in the x-z plane of the simulated glass for an isolated event in the elastic part of the strain curve at $\gamma = 0.025$. (A) in-plane displacements $(u_x,u_z)$(B) Burgers vector field $(b_x,b_z)$(C) quadrupolar field $(Q_{xx},Q_{xz})$ (D) particle displacements with Burgers vector direction overlaid. Here, original positions are shown by green particles and displaced particles are pink. Particle separations are 1 scaled unit.}
\label{fig5}
\end{figure}
To understand the spatial correlations between the different parameters in Figure~\ref{fig2} at a microscopic level we magnify the small, isolated plastic event in Figure~\ref{fig2} (A) and display the displacement, Burgers vector and quadrupole as vector fields alongside the particle positions before and after the strain increment in Figure~\ref{fig5}. Both the Burgers vector and the quadrupole show intricate patterns in the vicinity of the large displacement field. Interestingly, the quadrupolar field does not show the characteristic 4-fold symmetry of an Eshelby inclusion event (compare with Supplementary Material Figure~S.2). Also note that neither the quadrupolar or Burgers vector magnitudes are maximum in the exact area of maximum displacement magnitude. Although there are large displacements in this region, they have a low deviatoric component and are smoothly varying and continuous, giving rise to compatible strain.

The Burgers vector direction captures something essential about the coordinated slips in the dense system under strain as seen in Figure~\ref{fig5} (D) that overlays the Burgers direction on the particle snapshots before (green) and after (pink) the strain step. In the vicinity of this event the material undergoes coordinated slips in stripes that are angled at $45^{o}$ to the shear strain direction (which can be understood in a similar manner to Schmid's law in crystal plasticity \cite{baggioli}). These fine-grained stripes are the thickness of a polyhedral radius and reminiscent of nanometre-sized striped domains that have been found in some ferroelectrics \cite{wei2016neel} or strain-glass ferroelastic and shape-memory alloys \cite{XU2021117232,doi:10.1080/14786430903074771,PhysRevLett.95.205702}.

The centre of this isolated plastic event at (x,z) = (4,5) contains a particle with a large out-of-plane non-affine displacement. At a microscopic level, the correlation between large in-plane and out-of-plane displacements and large changes in local centrosymmetry can be understood intuitively. A polyhedron with a high centrosymmetry in the x-z plane will distort affinely with each applied strain increment as forces are balanced, and retain local centrosymmetry. However, if a particle jumps out of plane, many non-affine displacements can subsequently occur, resulting in a large change in the local centrosymmetry. This can lead to a cascade, or chain-reaction, effect in neighboring polyhedra. We see strong evidence for this in the mapped change in local centrosymmetry with coordinated positive/negative changes occurring in neighboring local structures. This observation may make a connection between defects and curvature that is very intuitive in 2D systems \cite{doi:10.1080/00018730903043166}, but less obvious in a dense 3D glass. Here defects form through particles jumping out of the plane defined by the applied strain direction. This mechanism is perhaps also related to bond-switching between neighboring tetrahedra that has been observed in a granular system under shear \cite{cao2018structural}. In general, displacements can only be measured directly in a glass made of particles that can be resolved optically, but changes in local distortions and centrosymmetry can be accessed using small-beam diffraction \cite{doi:10.1126/sciadv.abn0681}. In the following section, we assess how these measurable parameters can give insight into localized plastic events in a glass. 

\section{Continuous Burgers vector measurement from observable distortion}\label{sec3}
In a crystal, the reference lattice configuration is known from the atomic positions in the unit cell and lattice constants. Thus, even though displacements are needed to calculate the Burgers vector (Eqn.~\ref{burgers}), the Burgers vector in a crystal can be calculated using only the final distortions that are assumed to reflect the difference between the ``known" initial and final deformed structural configurations. This procedure has been used to find dislocations in crystals in both imaging and diffraction experiments using electrons and x-ray scattering \cite{HYTCH1998131,wheeler,cloete}. For a glass, the reference configuration is not known, which invites the question, is it possible to calculate the Burgers vector and quadrupolar strain for a glass without explicitly knowing the displacements? In geomechanics, strain from deformation events is inferred from the distribution of apparently randomly placed inclusions in a rock using the well-known Fry method to plot out the strain ellipse \cite{FRY197989}. Here, we test whether the same strategy can be employed for a glass.

\begin{figure}[!htbp]
\centering
\includegraphics[width=\linewidth]{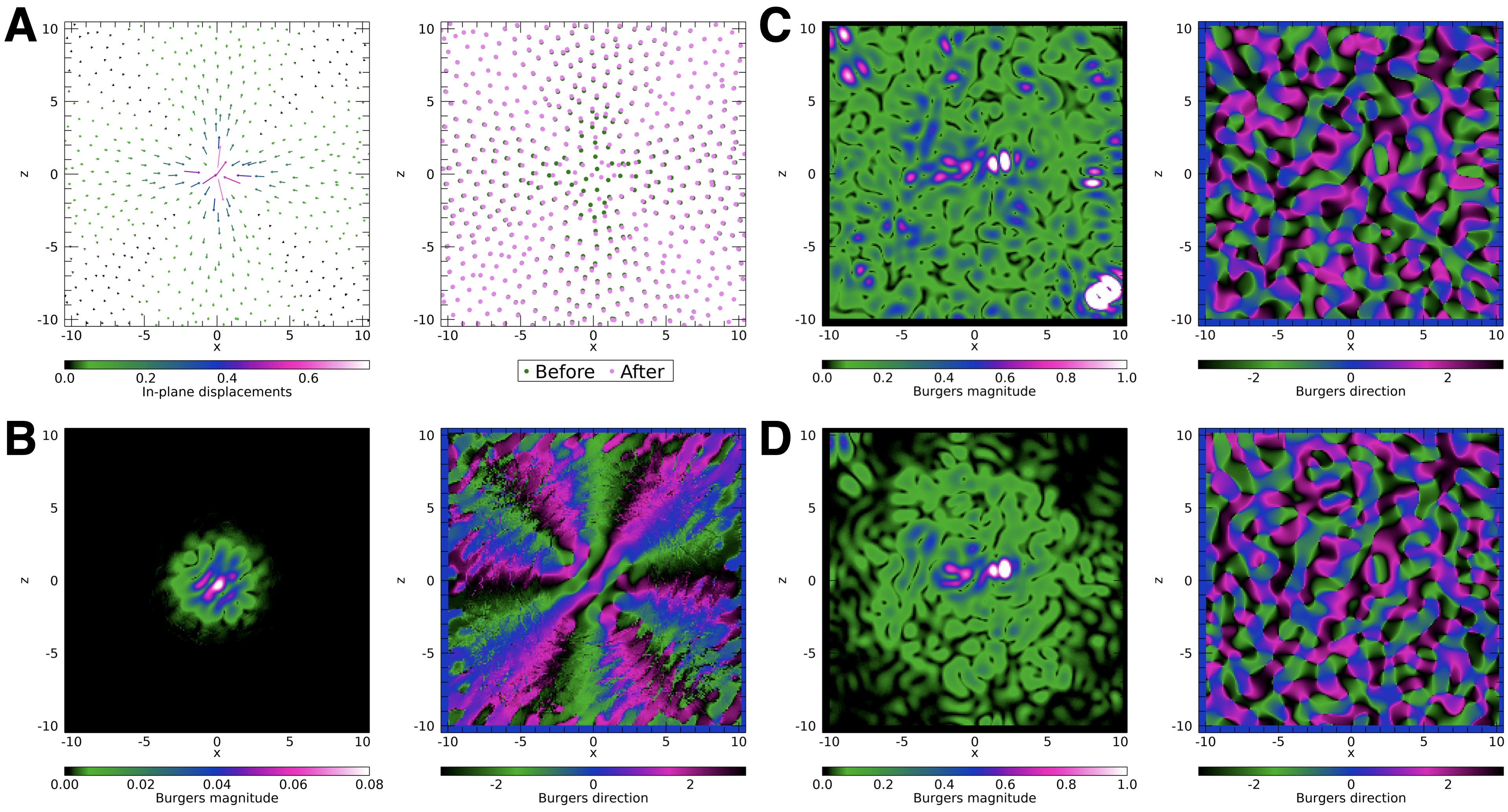}
\caption{Burgers vector from a localized quadrupolar event in a glass (A) Displacements and configuration Before (green) and After (Pink) local deformation (B) Burgers vector (magnitude and direction) calculated explicitly from displacements (C) Burgers vector (magnitude and direction) calculated from the distortion maps of the deformed configuration (D) Burgers vector (magnitude and direction) calculated from the final distortion with the initial distortion subtracted.}\label{fig6}
\end{figure}

In Figure~\ref{fig5} (A) we take a slice of our glass model, and introduce particle displacements that correspond to an ideal quadrupolar distortion field \cite{C7SM02510F}, namely:
\begin{equation}
    u_{r}(r,\phi) \propto \frac{\cos(2\phi)}{r} \textrm{ and }
    u_{\phi}(r,\phi) = 0,
\end{equation}
where $u_{r}$ is the radial component of the displacement and $u_{\phi}$ is the azimuthal component. The quadrupolar displacement field is clearly seen in the mapped displacement vectors but there are also clear changes in the local particle distribution in the configuration ``after" the localized deformation has been introduced (Figure~\ref{fig6} (A). If we calculate the Burgers vector and quadrupolar strain field from the displacements explicitly (Figure~\ref{fig6} (B) and Supplementary Material Figure~S.2 (B)) we uncover the magnitude and direction of the defect. If we fit the distribution of particle positions around each particle with a distortion tensor (see Supplementary Material Figure~S.4) from the configuration ``after" deformation, the localized quadrupolar event is still visible in the calculated Burgers vector and quadrupolar fields (Figure~\ref{fig6} (C) and Supplementary Material Figure~S.2 (C), respectively). But, it is of the same magnitude as the local distortions that pre-exist in the glass' configuration and form during solidification. This observation explains why the ideal quadrupolar Eshelby distortion may not be obvious in post-deformation particle-level strain fields of a glass \cite{doi:10.1126/sciadv.abn0681}. When averaging the strain over larger volumes and more extensive length scales (\textit{e.g.}, over many hundreds of local events) strain fields with quadrupolar symmetry clearly emerge due to the strong directionality of the strain fields and tendency for quadrupoles to align \cite{Wilde,pekin2019direct}. 

\textit{In situ} experiments can provide measurements of the system before and after straining \cite{pekin2019direct,doi:10.1126/sciadv.abn0681}. If we take ``before" and ``after" distortion maps, and subtract the initial distortions from the maps after deformation, we can recover the localized magnitude and azimuthal symmetry of the quadrupolar event in the Burgers vector and quadrupolar strain fields, albeit with some loss of clarity, especially for the direction (Figure~\ref{fig6} (D) and Supplementary Material Figure~S.2 (D), respectively). This discrepancy can be attributed to the accuracy of the fitted distortion and also the fact that for large displacements a particle may transfer from one nearest-neighbor polyhedron to another, giving an abrupt change in the fitted strain of those two polyhedra. In the limit of small displacements, the calculation of the Burgers vector from displacements and from the difference in the distortion before and after the deformation event should be the same (Methods~\ref{calcparam}). The loss of the Burgers direction information is exaggerated by noise and phase-wrapping where the displacement field approaches zero. We note that this same approach could be useful for heavily deformed or disordered crystals where the reference configuration is not known \textit{a priori}.

While both the Burgers vector and quadrupolar field seem to capture essential aspects of local deformation events in a glass, the Burgers vector may be preferable to calculate, for a number of reasons. Firstly, while the quadrupolar strain field finds events with a large deviatoric component and quadrupolar symmetry, the Burgers vector is perhaps more general, being sensitive to incompatible strains due to both displacements and rotations \cite{kleinert1989gauge}. This may prove more powerful to analyze the distribution of possible plastic structural transformations in a glass, that may not always adhere to the idealized symmetry of a particular defect like the Eshelby inclusion \cite{hentschel2023eshelby}. Glasses possess a spectrum of different local structures that are sometimes symmetric and sometimes distorted \cite{ROYALL20151}. At the microscopic level, ``inclusions" correspond to polyhedra with different local elastic responses that may not conform to the ideal ellipsoidal geometry. Additionally, the Burgers vector is insensitive to the addition of a smooth displacement field, as we show in the Supplementary Material Figure~S.3, whereas the addition of a smooth displacement field alters and obscures the quadrupolar strain field. This invariance of the Burgers vector to the addition of a smooth field illustrates its topological nature.

\section{Experimental observations of defects in a colloidal glass from diffraction}\label{sec4}
Here we show how these measurable parameters can be calculated from scanning small-beam diffraction experiments. The \textit{in situ} scanning microbeam small-angle x-ray measurement on colloidal glasses has been described previously \cite{doi:10.1126/sciadv.abn0681} and is analogous to scanning nanobeam electron diffraction that is readily implemented in a scanning/transmission electron microscope for atomic glasses \cite{PhysRevLett.110.205505}. Here the polyhedral diameter is also one length unit (in $\mu m$). We examine these spatially-resolved parameters from a colloidal glass specimen where in-plane shear strain has been applied through the withdrawal of a wire that was encapsulated inside the glass (see Supplementary Material Figure~S.5 and Methods~\ref{expt}).
\begin{figure}[!htbp]
\centering
\includegraphics[width=\linewidth]{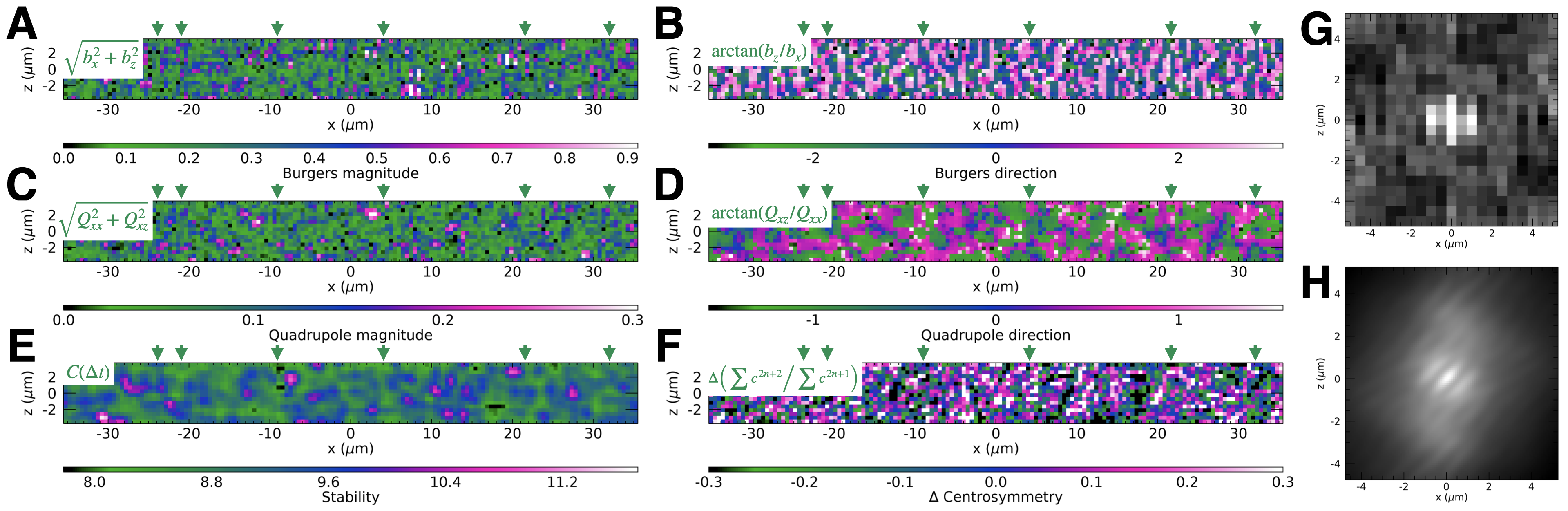}
\caption{Experimental maps from scanning microbeam small-angle x-ray scattering from a colloidal glass subjected to simple shear in the direction left-right (A) Burgers vector magnitude ($\sqrt{b_x^2+b_z^2}$) and (B) direction $\arctan{(b_z/b_x)}$. (C) Quadrupolar magnitude ($\sqrt{Q_{xx}^2+Q_{xz}^2}$) and (D) direction ($\arctan{(Q_{xz}/Q_{xx})}$) (E) Local stability ($C(\Delta t)$). (F) Change in local centrosymmetry ($\Delta \big(\sum c^{2n+2} \big/ \sum c^{2n+1}\big)$). (G) Autocorrelation of the experimental Burgers vector magnitude map (shown in (A)) showing stripes at the length scale of a polyhedral radius. (H) Autocorrelation of the simulated Burgers vector magnitude map (Figure.~\ref{fig2} (B)) showing striped texture at the same length scale, but with a more prominent texture direction of $45^{o}$ to the applied shear. Green arrows point to striped features at $45^{o}$ to the applied strain that match the features seen in the simulated sheared glass. Glasses were composed of 300~nm diameter colloidal SiO$_{2}$ particles dispersed in water with 0.1 M NaCl (~\ref{expt}).}\label{fig7}
\end{figure}
Figure~\ref{fig7} shows the Burgers vector magnitude and direction ((A) and (B)) from the measured final fitted local distortion with the initial fitted local distortion from the same region subtracted. Panels (C) and (D) display the quadrupolar magnitude and direction, again subtracting the initial strain. Panel (E) shows the measured local stability of the initial configuration (from time correlations in diffracted intensities in the microdiffraction patterns) and panel (F) shows the change in the local structural centrosymmetry from the breakdown in Friedel symmetry in the dynamical diffraction patterns \cite{doi:10.1126/sciadv.abn0681} (Methods~\ref{expt}). All these mapped parameters show spatial fluctuations on the length scale of individual nearest-neighbor polyhedra as expected from the simulations. Striped regions with large Burgers vector magnitude occur at an angle of $45^{o}$ to the applied strain direction (green arrows). The autocorrelation function of the Burgers vector magnitude (Figure~\ref{fig7} (G)) shows the same striped features at the same length scale as the simulation (Figure~\ref{fig7} (H) calculated from Figure.~\ref{fig2} (B)) even though the experimental sampling is ten times lower than the simulation. The Burgers vector direction and change in centrosymmetry maps show similar evidence of cooperative slip and structural transformations in stripes as seen in the simulated system. The Burgers vector and quadrupolar magnitudes show the same trends with centrosymmetry prior to deformation as observed in simulation as seen in Figure~\ref{fig3} (B); which is that lower centrosymmetry correlates with plasticity down to some threshold value. As observed previously, centrosymmetry correlates well to stability \cite{doi:10.1126/sciadv.abn0681}, although in this experimental study stability reflects particle displacements during relaxation. The inverse of the stability plot in Figure~\ref{fig3} (B) shows the same trends as the simulated displacement plot in Figure~\ref{fig3} (A). Beam stability, specimen stage precision and spatial sampling need to be improved to more thoroughly explore the relationships between all the parameters at the microscopic level.
\section*{Discussion and conclusion}\label{sec5}
We examine measurable geometric parameters that can be extracted from small-beam diffraction experiments to probe deformation-induced local structural transformations in glasses. This class of experiments could provide a useful, unified methodology to study deformation phenomena across the broad range of systems that possess disorder. In particular, they offer microscopic insight into plastic rearrangements in systems where the particle positions and displacements can't be spatially resolved in 3D, such as atomic glasses and colloidal systems with particles in the sub-micron range. We compare several local geometric parameters: the centrosymmetry in nearest-neighbors around a central particle, the continuous Burgers vector and the quadrupolar field that can both be derived from the local distortions. From these studies, we find that the local Burgers vector captures essential features of the magnitude and direction of the material slips in the system. The quadrupolar field often lacks the expected four-fold azimuthal symmetry, and moreover is highly sensitive to the addition of other smooth strain fields in contrast to the Burgers vector which is a topological invariant. The local centrosymmetry changes in a spatially coordinated way in the vicinity of plastic events. These changes in local structural symmetry seem to be connected to large out-of-plane particle displacements, which may link topological defects to particle motion perpendicular to the applied shear plane in a glass. 

Our study suggests many further experimental and theoretical directions. We examined colloidal glasses composed of microspheres with x-ray microbeam diffraction, but an analogous experiment in the scanning/transmission electron microscope using electron nanodiffraction can be employed to study deformation-induced local structural transformations in atomic glasses \cite{PhysRevLett.110.205505,pekin2019direct}. Here we took a quasi-2D approach by slicing the 3D sample to compare to transmission diffraction measurements on thin (approximately 5-10 polyhedral diameters thick) glass specimens where the structure is projected \cite{doi:10.1126/sciadv.abn0681,PhysRevLett.110.205505}. However, this analysis can be extended into the third dimension for more bulk specimens with diffraction-based strain tomography \cite{lionheart2015diffraction,tovey2020scanning} and Burgers vector analysis in 3D \cite{cloete}. Such an extension could be used to visualize and quantify the incompatible strains that are quenched into a glass during formation, evolve during ageing and arise abruptly during plastic deformation, a significant new tool for glass science and engineering. The Burgers vector is a powerful analytical tool that could shed new light on the nature of the defects that carry plasticity in an amorphous material. The observation of coordinated slips and changes in structural symmetry at the length scale of a polyhedral radius coincides with the length scale for softness identified from machine-learning in many systems \cite{doi:10.1126/science.aai8830}. These results give microscopic insight into how structure may relate to stability and plasticity in a glass and could provide new directions to engineer stable crystals and glasses for different applications.

\section{Methods}\label{sec6}
\subsection{Modeling and simulation}\label{model}
Kremer-Grest models were employed \cite{PhysRevA.33.3628} consisting of 50-monomer linear chains. The polymer chain had two different types of masses ($m_{1} = 1$ and $m_{2} = 3$ placed in alternating order). The total number of monomers in the system was N = 10000. Glasses initially quenched to zero in temperature are deformed using the athermal quasi-static shear protocol (AQS) in the LAMMPS simulation package \cite{PLIMPTON19951} using increments in simple shear of $\delta \gamma_{xz} = 0.001$. The configuration is saved after each strain and relaxation step. Further details of modeling can be found in \cite{PhysRevB.102.024108}.

\subsection{Calculation of parameters}\label{calcparam}
The models were sliced into twenty slices that were $2R_{0}$ in thickness where $R_{0}=1$ was the average inter-particle separation of 1. This quasi-2D treatment was performed in the x-z direction at every strain step and parameters were calculated and averaged to compare to the stress-strain curve. Higher resolution volumetric calculations were performed at strains corresponding to different elastic/plastic events by slicing the models in 270 slices using a rolling window that was $2R_{0}$ in thickness.

The degree of inversion symmetry or centrosymmetry in nearest-neighbor particle arrangements was calculated according to Eqn.~\ref{centro} and employing a radial cutoff in neighbors of $1.2$. This field was interpolated on a regular grid (270x270) using the GRIDDATA function in Interactive Data Language (IDL) and a radial basis function.

In Section~\ref{sec2} the distortion tensor was derived from the known particle displacements. The non-affine displacements of the particles in a slice ($u_{i}^{NA}$) were calculated by subtracting the affine component of the applied displacements ($u_{i}^{A}$) as follows: $u_{i}^{NA} = u_{i} - u_{i}^{A}$. These non-affine displacements were interpolated onto the same grid as the inversion symmetry parameter. The interpolated non-affine displacements were employed to calculate the components of the in-plane distortion tensor using finite differences. The 2D distortion tensor is given by \cite{nyebook}:
\begin{equation}
 e_{ij} = \begin{pmatrix}
\frac{\partial u_{1}}{\partial x_{1}} & \frac{\partial u_{1}}{\partial x_{2}} \\
\frac{\partial u_{2}}{\partial x_{1}} & \frac{\partial u_{2}}{\partial x_{2}}
\end{pmatrix}
= \begin{pmatrix}
e_{11} & e_{12} \\
e_{21} & e_{22}
\end{pmatrix} = \epsilon_{ij} + w_{ij}
\end{equation}
Here $\epsilon_{ij}$ is the symmetric component of the strain:
\begin{equation}
    \epsilon_{ij} = \frac{1}{2}(e_{ij} + e_{ji}),
\end{equation}
and $w_{ij}$ is the antisymmetric part:
\begin{equation}
    w_{ij} = \frac{1}{2}(e_{ij} - e_{ji}),
\end{equation}
that break the distortion down into two components: strains (volumetric and shear) and rotations, respectively.

The Burgers vector was calculated using the distortion tensor components as follows:
 \begin{equation}
    b_{i} = -  \oint\, du_{i} =  -\oint\, \frac{\partial u_{i}}{\partial x_{j}} dx_{j} = -\oint\, e_{ij} \frac{dx_{j}}{dp}dp,
    \label{burgers1}
\end{equation}
The contour integral to calculate the local Burgers vector numerically was a square 7x7 pixels in size employing four line segments parameterized by the variable $p$, as described in \cite{baggioli}. Here for a square with side length $2L$ and a centre of $(x_{1}(0),x_{2}(0))$, four line segments A,B,C,D going anticlockwise around the contour are defined by:

\begin{equation*}
  l_{A}:\begin{cases}
    x_{1}(p) &= x_{1}(0) - L + p \\
    x_{2}(p) &= x_{2}(0) - L 
  \end{cases} \\
\end{equation*}
\begin{equation*}
  l_{B}:\begin{cases}
  x_{1}(p) &= x_{1}(0) + L \\
  x_{2}(p) &= x_{2}(0) - L + p 
    \end{cases}
\end{equation*}
\begin{equation}
  l_{C}:\begin{cases}
    x_{1}(p) &= x_{1}(0) + L - p \\
    x_{2}(p) &= x_{2}(0) + L 
  \end{cases} \\
\end{equation}
\begin{equation*}
  l_{D}:\begin{cases}
  x_{1}(p) &= x_{1}(0) - L \\
  x_{2}(p) &= x_{2}(0) + L - p 
    \end{cases}
\end{equation*}
for $p \in [0,2L]$.

 The traceless symmetric strains from Eqn.~\ref{quad} were used to calculate the quadrupolar field. The magnitude and direction of the quadrupolar field are given by:
\begin{equation}
    Q_{mag}^{2} = Q_{11}^{2} + Q_{12}^{2}  ,
    \label{quadm}
\end{equation}
\begin{equation}
    \theta = \frac{1}{2} \mathrm{arctan} (Q_{12}/Q_{11}) , 
    \label{quadd}
\end{equation}
that reflect the definition of the quadrupolar component as \cite{Kumar_2024,PhysRevE.104.024904}:
\begin{equation}
    Q_{ij}=\begin{pmatrix} Q_{11}~Q_{12}\\
Q_{21}~Q_{11} \end{pmatrix}=Q_{mag}\begin{pmatrix} \cos{2\theta}~\sin{2\theta}\\
\sin{2\theta}~-\cos{2\theta} \end{pmatrix}.
\end{equation}

Volumetric renders of the high resolution parameters calculated from the simulated system were prepared in Napari \cite{sofroniew_2023_8076424}. The local structural centrosymmetry, $F_{IS}$, was displayed as an isosurface with a value of 0.75 and an opacity of 0.36. The non-affine displacement magnitudes were displayed with the following contrast ranges: $\gamma = 0.025$, 0-0.0043; $\gamma = 0.093$, 0-0.0112; $\gamma = 0.108$, 0-0.7. The non-affine displacement magnitudes were rendered as an attenuated maximum intensity projection with an opacity of 1.0.

In Sections~\ref{sec3} and ~\ref{sec4} the distortion tensor was derived from nearest-neighbor configurations and small-beam diffraction patterns that probe the nearest-neighbor shell ``before" and ``after" deformation was applied. In the simulated system, nearest-neighbors were identified using a radial cutoff of $R_{max}=1.2$. The local distortion in nearest-neighbor configurations was estimated by fitting the positions of the neighbors with a distortion tensor:
 \begin{equation}
    R_{\alpha \beta} = e_{ij} R_{0} \hat{n}_{\alpha \beta} 
    \label{fit}
\end{equation}
Here $R_{\alpha \beta}$ is the vector between a central particle and a particle in the nearest-neighbor shell, $R_{0}$ is the average neighbor distance and $\hat{n}_{\alpha \beta}$ is a unit vector pointing from $\alpha$ to $\beta$. Examples of fits from a nearest-neighbor configuration ``before" and ``after" deformation are shown in Figure~S.4 (A) and (B), respectively.

The distortion introduced by the deformation increment was obtained from $e_{ij}^{d} = e_{ij}^{a} - e_{ij}^{b}$, where $b$ denotes ``before" and $a$ denotes ``after". As $\frac{\partial u^{a}_{i}}{\partial x_{j}} = \frac{\partial}{\partial x_{j}} \left(u^{b}_{i} + u^{d}_{i}\right) = \frac{\partial u^{b}_{i}}{\partial x_{j}} + \frac{\partial u^{d}_{i}}{\partial x_{j}}$, this is the same as the distortion derived explicitly from the displacements up to the precision of the fit, and the retention of particles in the same nearest-neighbor configuration. The derived distortion fields were interpolated on regular grids with the same spatial sampling as the non-affine displacements.

\subsection{Experimental scanning small-angle x-ray scattering}\label{expt}
Further details of colloidal dispersion preparation and x-ray experimental conditions have been reported previously \cite{aliu2017,doi:10.1126/sciadv.abn0681}. 300~nm diameter colloidal SiO$_{2}$ particles (polydispersity 5\%, Bangs Labs, IN, USA) were prepared as 2.5\% dispersions in water with 0.1 M NaCl. The dispersions were centrifuged into glasses for 10 minutes at 10,000 RPM in an Eppendorf miniSpin and immediately encapsulated between two $15 \mu$m Kapton films in the specimen holder with a spacer layer to control the thickness to $20 \mu$m and sealed. A $20 \mu$m diameter copper wire was embedded in the colloidal glass and drawn out of the glass at a rate of approximately 0.1~mm/s and a distance of 0.2 mm, producing a simple shear strain that varied linearly with distance from the wire (see Figure~S.5 (B)). Areas in the glass were scanned with the x-ray micro-beam before and after the shear strain was applied. 

$\mu$SAXS patterns were obtained by scanning the specimen with an ultra-low-divergence 5.5~keV x-ray micro-beam defined by a gold near-field aperture ($1.2~\mu$m diameter see Figure~S.5 (A)) milled into a $25~\mu$m thick gold foil with a focussed ion beam (Thermofisher Quanta 3D, 30~keV, 1.5 pA Ga$^{+}$ ion beam).  This aperture had smooth edges, giving rise to a well-defined Airy diffraction pattern (Figure~S.6 (A)). The small volume probed by the x-ray micro-beam produces well-resolved intensity ``speckles" in the diffraction pattern \cite{aliu2017,doi:10.1126/sciadv.abn0681}. Specimen scans were conducted in 0.5~$\mu$m steps and each line was scanned twice to collect sets of $I(q_{x},q_{z},x,z, t)$ and $I(q_{x},q_{z},x,z, t+\Delta t)$ with $\Delta t$ = 30~s.  Diffraction patterns were collected using a Dectris - Pilatus 2M direct detection camera at a camera length of 7244 mm. Examples are displayed in Figure~\ref{fig1a}. These diffraction patterns display prominent diffraction speckle due to the colloidal glass and also finer scale ($\approx 0.0001 \AA^{-1}$), radial features that may correspond to the Kapton films behaving like a Fabry-P\'{e}rot etalon \cite{etalon1,etalon2,etalon3}. These fringes confirm the coherence of the beam formed by the small aperture. Fringes like these were not observed in the diffraction patterns obtained with a much larger aperture \cite{doi:10.1126/sciadv.abn0681,aliu2017}, showing the coherence of this $1.2~\mu$m diameter beam. The analysis of the speckle from the colloidal glass is robust in the presence of these fine fringes as each analysis method averages over a q-range ten times bigger than the fringe spacing.

To quantify subtle angular symmetries in the ``speckle" diffraction patterns the angular autocorrelation function ($C(q,\Delta)$) of each diffraction pattern ($I(q_{x},q_{z},x,z)$ = $I(q, \phi)$) was calculated: 
\begin{equation}
C(q,\Delta) = \frac{\langle I(q, \phi)I(q, \phi + \Delta) \rangle_{\phi} - \langle I(q, \phi) \rangle_{\phi}^{2}}{\langle I(q, \phi) \rangle_{\phi}^{2}} .
\label{qcor}
\end{equation}

Here the scattering vector magnitude is $|\vec{q}| = q = \sqrt{q_{x}^{2}+q_{z}^{2}} = 4\pi \mbox{sin} \theta/\lambda$ with a scattering angle of $\theta$ and x-ray wavelength of $\lambda$. $\phi$ is the azimuthal angle in the diffraction plane and $\langle X \rangle_{\phi} = \frac{1}{2 \pi} \int_{0}^{2 \pi} X d\phi$ denotes averaging over $\phi$. Normalized symmetry magnitudes $c_{q}^{n}\slash c_{q}^{0}$ were calculated from the Fourier transform of $C(q,\Delta)$ \cite{aliu2016b} at each value of $q$.

The appearance of non-zero odd-order Fourier coefficients indicates a break-down of $\pi$-rotational symmetry in the diffraction pattern. As demonstrated in previous simulation studies the lack of this inversion or Friedel symmetry in the diffraction pattern in this scattering geometry can be attributed to dynamical diffraction effects and an object that is non-centrosymmetric in the specimen plane \cite{doi:10.1126/sciadv.abn0681}. For thin specimens ($20 \mu$m), examining $q$-ranges in the first diffracted ring and using a micro-beam defined by a near-field aperture, Ewald sphere curvature and aberrations in the incident beam are not expected to contribute to Friedel symmetry breaking \cite{acyliu2015,doi:10.1126/sciadv.abn0681}. Thus to map local centrosymmetry, the parameter $\sum c^{2n+2}\slash \sum c^{2n+1}$ for $0 \le n \le 5$ was recently proposed as it can be measured directly from small-beam diffraction patterns \cite{doi:10.1126/sciadv.abn0681}. This parameter probes the relative degree of local centrosymmetry in particle arrangements within the near-field aperture. 

The local time correlation coefficient at each point in the scanned array $(x,z)$ was calculated from the intensity ($I(q_{x},q_{z},x,z, t)$ = $I(t)$) according to:
\begin{equation}
C(\Delta t) = \frac{\langle (I(t) - \langle I(t)\rangle_{q1})(I(t + \Delta t) - \langle I(t + \Delta t)\rangle_{q1}\rangle_{q1}}{\langle I(t) \rangle_{q1} \langle I(t + \Delta t) \rangle_{q1}} .
\label{tcor}
\end{equation}

This local time correlation coefficient varies from point-to-point enabling the heterogeneity in local structural dynamics in the glass to be visualized.
          
The local distortion tensor was calculated from experimental diffraction patterns by dividing the diffracted intensity into angular ranges of $\pi/24$ and averaging the diffracted intensity over this angular range to get intensity profiles, $I_{\phi}(q)$.  The intensity centre-of-mass from each arc, $\sum_{n} q_{n} I_{\phi}(q_{n})$, was calculated to give the value of $q_{max}$ where the peak in the diffracted intensity was located.  $q_{max}$ was normalized by the average $q_{0} = \langle q_{max} \rangle_{\phi}$ from the average $\mu$SAXS pattern of the entire ensemble.  The normalized $q_{max}$ was fitted with a distortion tensor in the plane of the specimen \cite{doi:10.1126/sciadv.abn0681,pekin2019direct} :
\begin{equation}
\frac{q_{0} - q_{max}(\phi)}{q_{max}(\phi)} = \frac{ R_{\phi}\hat{\phi}}{R_{0}} = \frac{e_{ij} R_{0}\hat{\phi}}{R_{0}} = e_{ij}\hat{\phi} \\
\label{anis}
\end{equation}
with $\hat{\phi} = (\cos(\phi), \sin(\phi))$. $q_{0}$ was determined from the averaged diffraction pattern for the glass prior to deformation. Examples of fits from the same volume in the glass ``before" and ``after" deformation are shown in Figure~S.6 (C) and (D), respectively. Further more extensive investigations of this process are required to identify the optimum procedure and parameters for this fit and limits of sensitivity (\textit{e.g.} number of points, infinitesimal or finite deformation etc.)  \cite{villert2009,BRITTON201282}.

The parameters derived from the distortion tensor - the Burgers vector and quadrupolar field - were calculated as above for the simulations, taking care to subtract the distortion field from the initial configuration. A 3x3 closed contour integral was used to determine the Burgers vector, due to the low sampling for the experimental data. We note that the value of $q_{0}$ does not affect the calculation of the Burgers vector, but the quadrupolar field is sensitive to the choice of the average volumetric strain. 

Experimental maps were smoothed for display using a Savitsky-Golay smoothing kernel of width 2 and order 1. This optimal least-squares fitting method smooths pixel-wise noise, whilst retaining the width and magnitude of peaks \cite{doi:10.1126/sciadv.abn0681}. Other parameters like the autocorrelation and average values corresponding to different quartiles in the parameter distributions were calculated from the raw maps with no filtering applied. The autocorrelation function is calculated from a zero-padded array of twice the largest array dimension as follows:

\begin{equation}
\mathrm{Corr}(\Delta x,\Delta z) = \frac{\langle [f(x,z)-\langle f(x,z) \rangle_{x,z}] [f(x+\Delta x,z+\Delta z)-\langle f(x,z) \rangle_{x,z}] \rangle_{x,z}}{\sigma_{f(x,z)}\sigma_{f(x,z)}}.
\label{ccor}
\end{equation}
where $\langle f(x,z) \rangle_{x,z}$ is the average value of the array and $\sigma_{f(x,z)}$ is the standard deviation.

\backmatter

\bmhead{Supplementary information}
Supplementary information is provided.

\bmhead{Acknowledgements}
M.B. thanks W.~Kob, Z.~Wu, J.~Zhang, Y.~Wang, C.~Jiang, Z.~Zheng, H.~Tong for useful discussions. A.L. thanks T. Finlayson and T. J. Davis for stimulating discussions.  
\section*{Declarations}

\begin{itemize}
\item MB acknowledges the support of the Shanghai Municipal Science and Technology Major Project (Grant No.2019SHZDZX01) and sponsorship from the Yangyang Development Fund. ACYL, TP and HP acknowledge support from the Australian Research Council (FT180100594, DP250102966). AZ gratefully acknowledges funding from the European Union through Horizon Europe ERC Grant number: 101043968 ``Multimech'', and from the Nieders{\"a}chsische Akademie der Wissenschaften zu G{\"o}ttingen in the frame of the Gauss Professorship program. AZ and AB gratefully acknowledge funding from US Army Research Office through contract nr. W911NF-22-2-0256. JNI acknowledges support from the Chan Zuckerberg Initiative (CZI) (grants SVCF 2018-192630, 2021-238853, 2023-332000). The authors acknowledge the use of the instruments and scientific and technical assistance at the Monash Centre for Electron Microscopy, a Node of Microscopy Australia. This research used equipment funded by Australian Research Council grant (LE0882821). This research was undertaken on the SAXS/WAXS beamline at the Australian Synchrotron, part of ANSTO.

\item The authors declare no competing interests.
\item Ethics approval and consent to participate. 
Not applicable.
\item Consent for publication
All authors consent.
\item Data and polymer glass models are available at the Monash Figshare repository: https://doi.org/10.26180/27193851.v1 and https://doi.org/10.26180/26199917.v1, respectively.
\item Materials availability
Not applicable.
\item Code availability
Code will be made available upon request to the authors.
\item Author contribution
ACYL/RFT/TCP/STM conceived and conducted the experiment. TWS simulated the model polymer glass. ACYL/HP/MB/AB/AZ conceived the analysis approach. ACYL/JN-I performed the analysis and data presentation. All authors prepared and reviewed the manuscript and Supplementary Material.   
\end{itemize}

\bibliography{bibliography}
\newpage

\begin{appendices}
\section*{Supplementary Material}\label{secA1}
\renewcommand{\thefigure}{S.\arabic{figure}}
\begin{figure}[h]
\centering
\includegraphics[width=\linewidth]{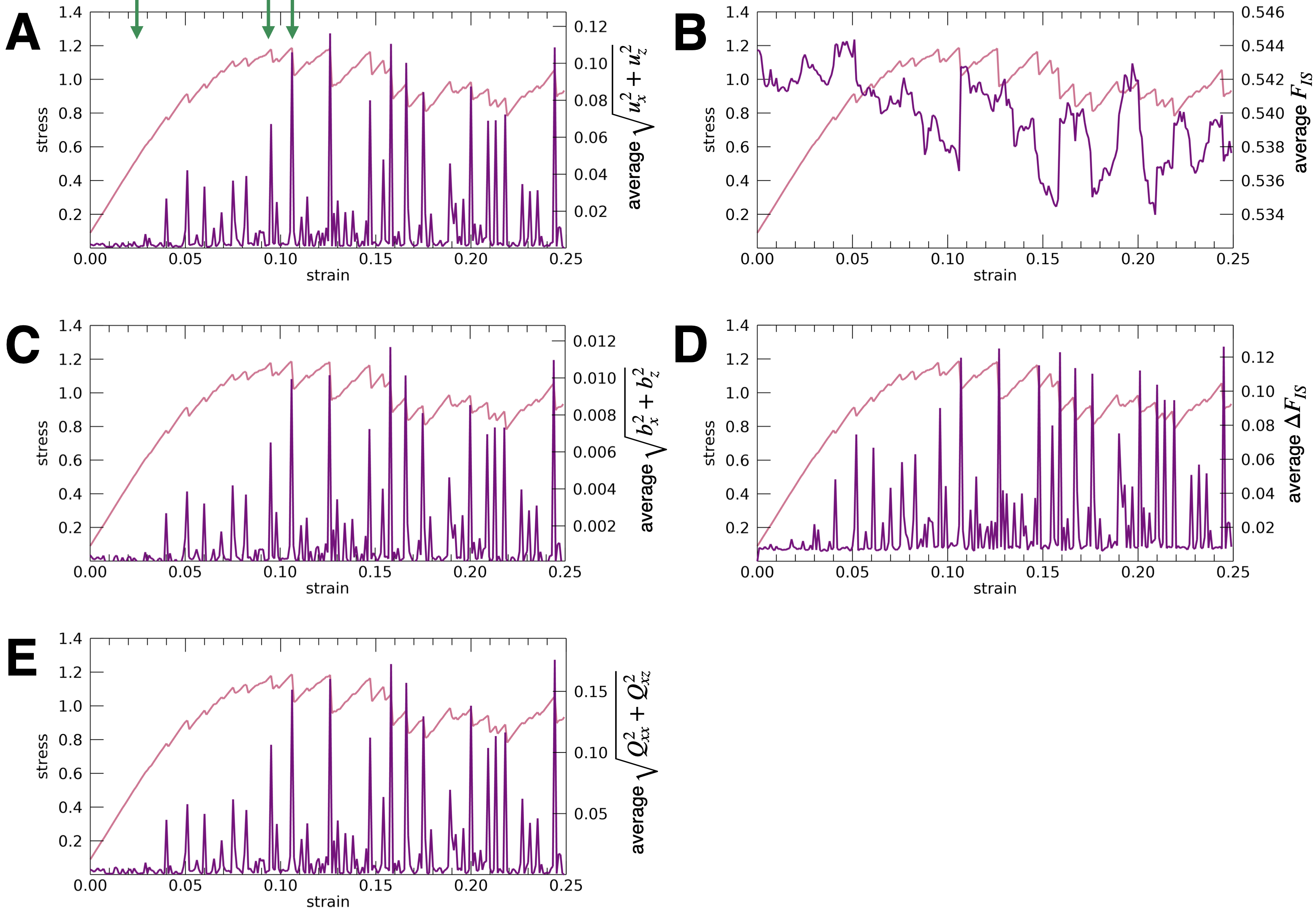}
\caption{Stress-strain curve (pink lines) from the simulated glass from the range of strain values studied $0.00 \leq \gamma \leq 0.25$. Green arrows show the values of strain for the configurations studied in Figure~\ref{fig2} of the main text. Overlaid on the stress strain curves are average parameter values from the x-z slices at the different values of strain. (A) Average displacement magnitude ($\sqrt{u_{x}^2+u_{z}^2}$), (C) Burgers vector magnitude ($\sqrt{b_{x}^2+b_{z}^2}$), (E) quadrupole magnitude ($\sqrt{Q_{xx}^2+Q_{xz}^2}$), (B) magnitude of the local centrosymmetry ($F_{IS}$)and (D) magnitude of the  change in local centrosymmetry ($\Delta F_{IS}$).}
\label{sfig1}
\end{figure}

\begin{figure}[h]
\centering
\includegraphics[width=\linewidth]{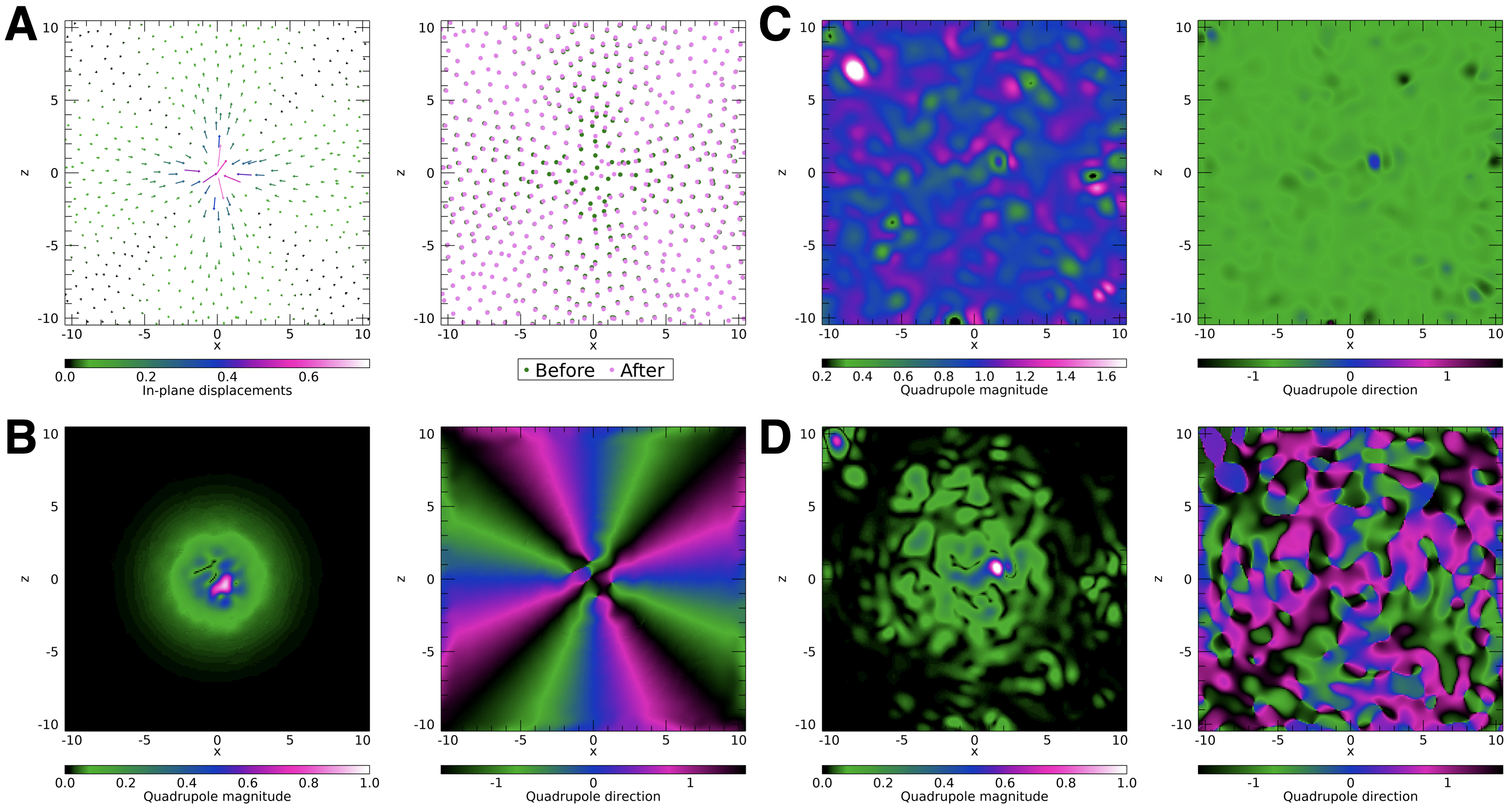}
\caption{Quadrupolar field from a localized quadrupolar event in a glass (A) Displacements and configuration Before (green) and After (Pink) local deformation (B) Quadrupolar field (magnitude and direction) calculated explicitly from displacements (C) Quadrupolar field (magnitude and direction) calculated from the strain maps of the deformed configuration (D) Quadrupolar field (magnitude and direction) calculated from the final strain with the initial strain subtracted. }
\label{sfig2}
\end{figure}

\begin{figure}[h]
\centering
\includegraphics[width=\linewidth]{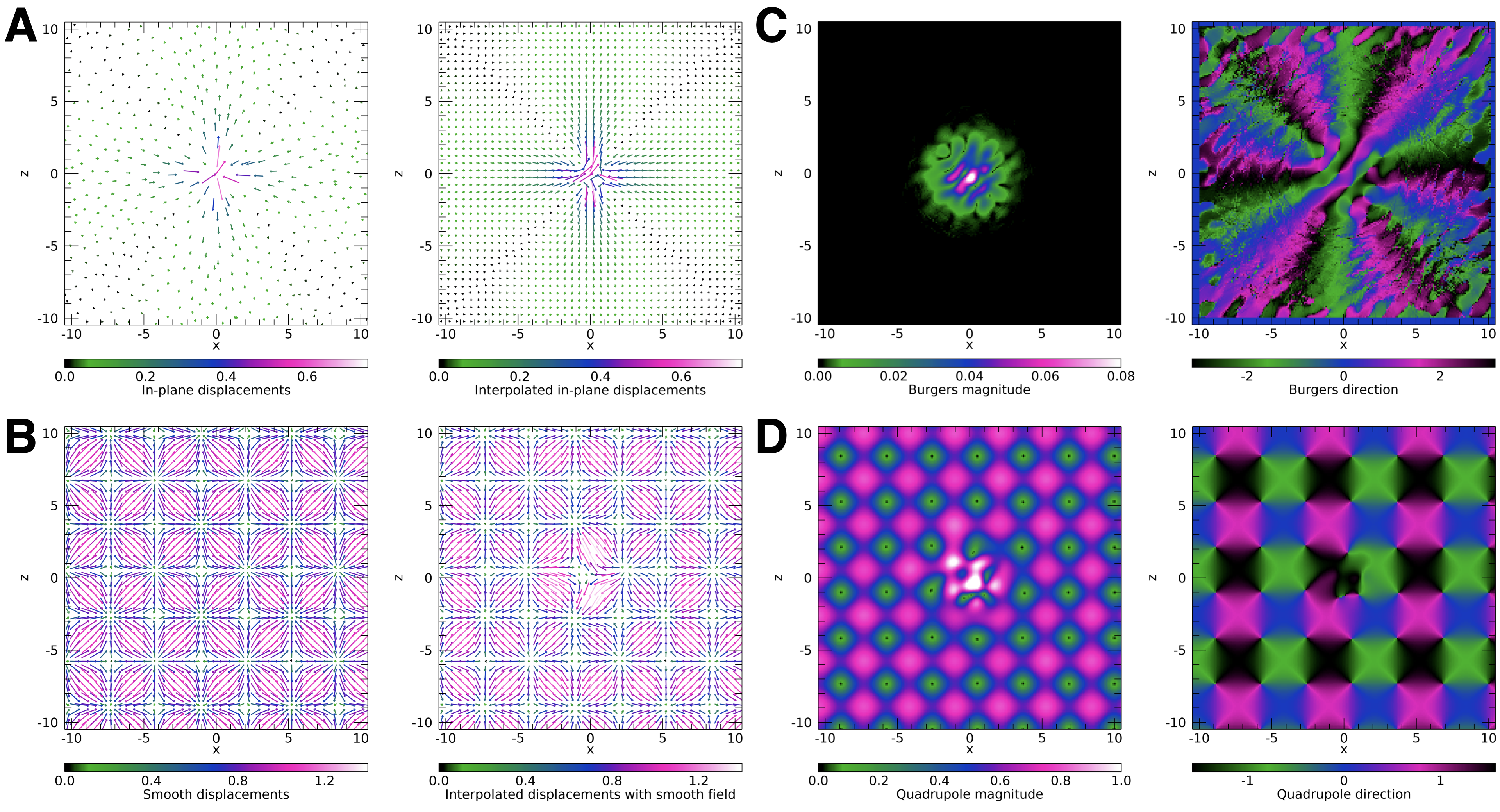}
\caption{Burgers vector and quadrupolar field  from a localized quadrupolar event in a glass with the addition of a smooth displacement field (A) Displacements from the localized quadrupolar deformation (B) Left - a smooth displacement field composed of sinusoids Right - displacements in (A) with the addition of the smooth field (C) Burgers vector (magnitude and direction) calculated from displacements in (B) that is insensitive to the addition of the smooth field (D) Quadrupolar field (magnitude and direction) calculated from the displacements that is heavily distorted by the addition of the smooth field.}
\label{sfig3}
\end{figure}

\begin{figure}[h]
\centering
\includegraphics[width=\linewidth]{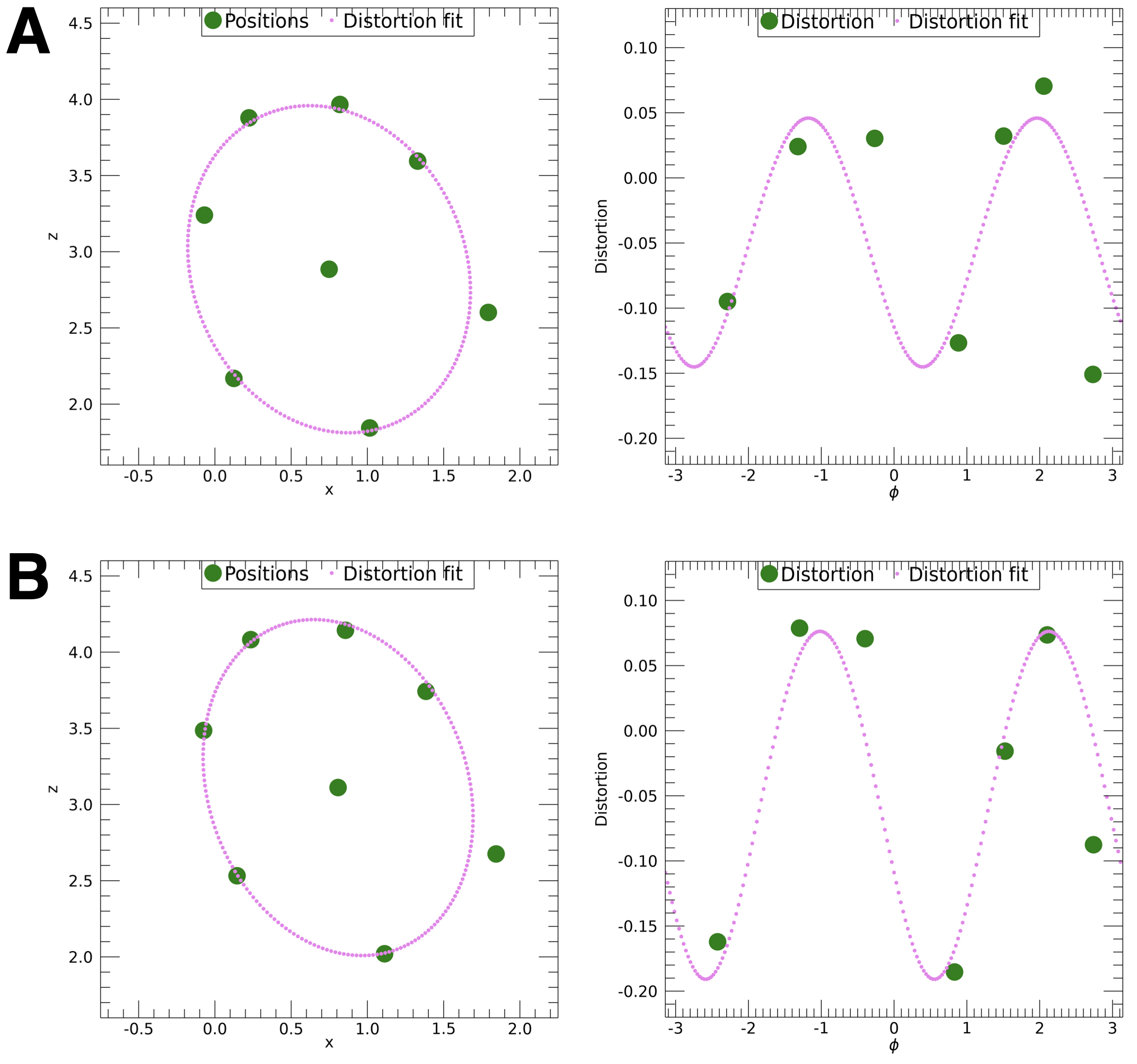}
\caption{Estimation of the distortion tensor employing snapshots of a nearest-neighbor configuration from simulation (A) ``before" and (B) ``after" deformation. The left panel shows the fit as an ellipse in the x-z plane, while the right panel shows the fit as a distortion as a function of azimuthal angle, $\phi$. }
\label{sfig4}
\end{figure}

\begin{figure}[h]
\centering
\includegraphics[width=\linewidth]{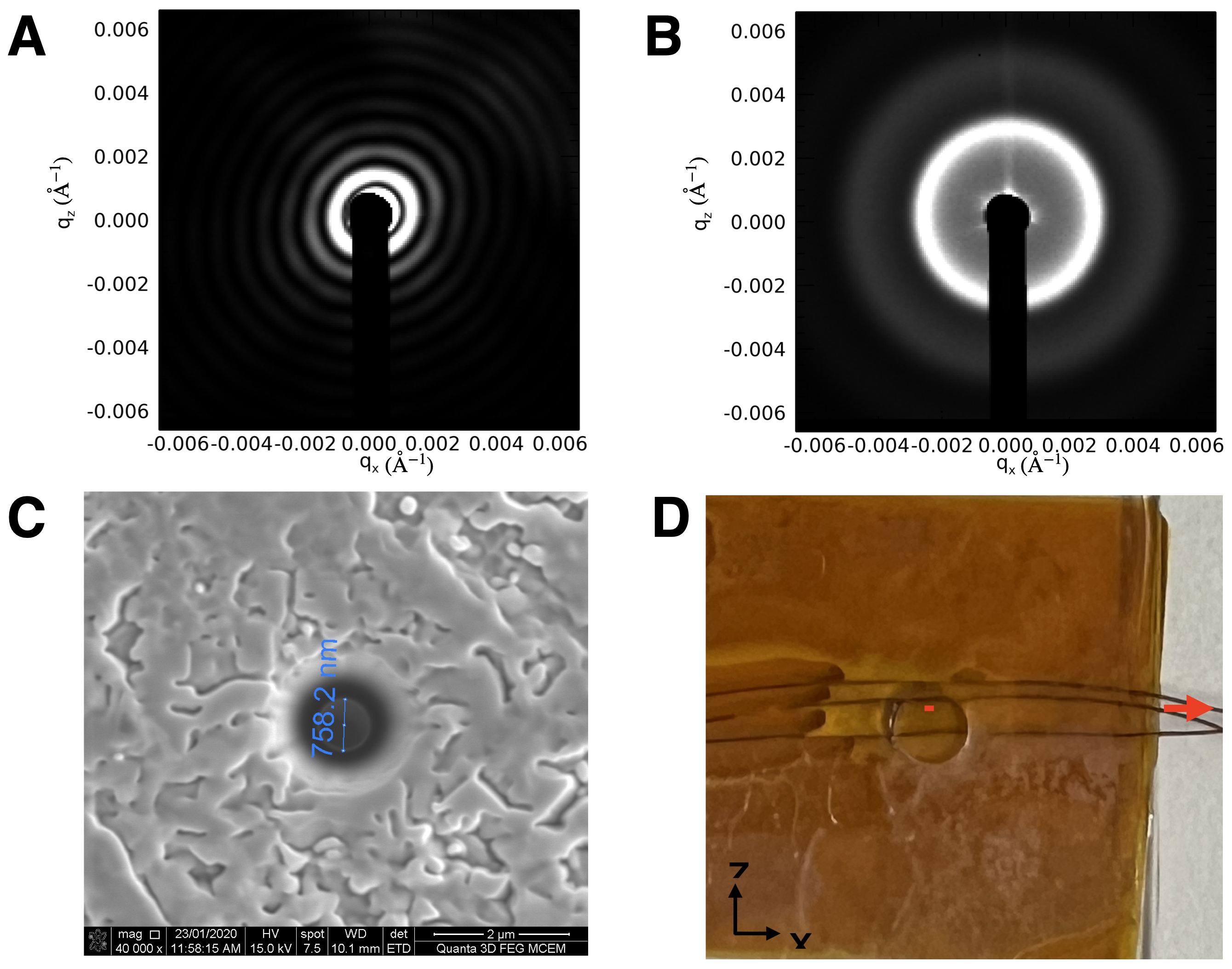}
\caption{(A) SAXS pattern of the $1 \mu$~m near-field aperture showing symmetric and smooth Airy rings. (B) Broad-beam SAXS pattern of the colloidal glass. (C) $1.2~\mu$m diameter near-field aperture (D) Colloidal glass specimen encapsulated between Kapton sheets with wires embedded to provide a linear shear profile in the plane of the glass. The red line below the centre wire shows the scanned area. }
\label{sfig5}
\end{figure}

\end{appendices}
\end{document}